\documentclass[pdflatex, sn-basic]{sn-jnl}

\usepackage{graphicx}%
\usepackage{multirow}%
\usepackage{amsmath,amssymb,amsfonts}%
\usepackage{amsthm}%
\usepackage{mathrsfs}%
\usepackage[title]{appendix}%
\usepackage{xcolor}%
\usepackage{textcomp}%
\usepackage{manyfoot}%
\usepackage{booktabs}%
\usepackage{algorithm}%
\usepackage{algorithmicx}%
\usepackage{algpseudocode}%
\usepackage{listings}%
\usepackage{subcaption}
\usepackage{url}
\definecolor{newcolor}{rgb}{.8,.349,.1}
\usepackage{tikz}	
\usepackage{hyperref}
\usepackage[T1]{fontenc}
\usepackage{latexsym}
\usepackage{verbatim}
\usepackage[section]{placeins}
\usepackage{xurl}

\theoremstyle{thmstyleone}%
\theoremstyle{thmstyletwo}%

\theoremstyle{thmstylethree}%

\begin{document}

\title[Article Title]{Towards CPU Performance Prediction: New Challenge Benchmark Dataset and Novel Approach}


\author*[1]{\fnm{Xiaoman} \sur{Liu}}\email{xiaoman.liu@intel.com}
\affil*[1]{\orgdiv{DCAI}, \orgname{Intel China Ltd.}, \orgaddress{\street{No. 880 Zi Xing Road}, \city{Shanghai}, \postcode{200241}, \country{China}} \\
\href{https://xiaoman-liu.com/}{Personal Homepage}
}


\abstract{

The server central processing unit (CPU) market continues to exhibit robust demand due to the rising global need for computing power. Against this backdrop, CPU benchmark performance prediction is crucial for architecture designers. It offers profound insights for optimizing system designs and significantly reduces the time required for benchmark testing.  However, the current research suffers from a lack of a unified, standard and a comprehensive dataset covering various CPU benchmark suites on real machines. Additionally, the traditional simulation-based methods suffer from slow simulation speeds. Furthermore, traditional machine learning approaches not only struggle to process complex features across various hardware configurations but also fall short in achieving sufficient accuracy.

To bridge these gaps, we firstly perform a streamlined data preprocessing and reorganize our in-house datasets gathered from a variety CPU models of 4th Generation Intel\textsuperscript{\textregistered} Xeon\textsuperscript{\textregistered} Scalable Processors on various benchmark suites. We then propose Nova CPU Performance Predictor (NCPP), a deep learning model with attention mechanisms, specifically designed to predict CPU performance across various benchmarks. Our model effectively captures key hardware configurations affecting performance in across various benchmarks. Moreover, we compare eight mainstream machine learning methods, demonstrating the significant advantages of our model in terms of accuracy and explainability over existing approaches.  Finally, our results provide new perspectives and practical strategies for hardware designers. To foster further research and collaboration, we \textit{\textbf{open-source}} the model \url{https://github.com/xiaoman-liu/NCPP}.

}




\maketitle
\section{Introduction}

\begin{figure}
	\centering
	\includegraphics[width=0.8\textwidth]{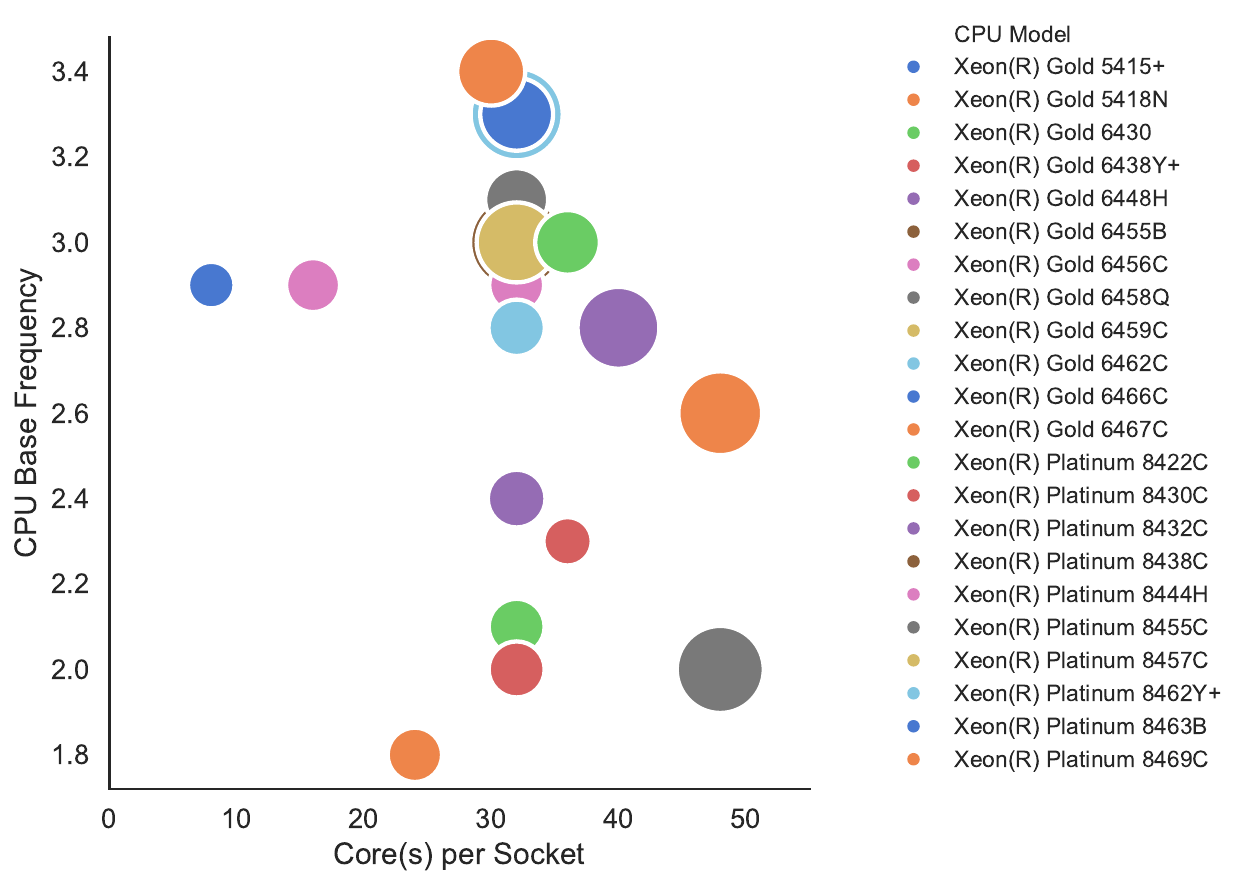}
	\caption{Bubble Chart Comparison of Core Count and Base Frequency Distribution for Various Intel Xeon Server CPU Models, with Bubble Size Representing Sample Size}
	\label{fig:product}
\end{figure}

As the brain of a computer system, the CPU handles the assignment and processing of tasks, and manages operational functions used by all types of computers. In recent years, the explosive surge in the demand for computational resources from both academia and industry has led to increasingly stringent requirements for high-performance CPU. Figure \ref{fig:product} presents the numerous CPUs that have been newly designed and manufactured in recent years. However, the architecture of CPU has become increasingly complex, presenting significant challenges to CPU performance evaluation. 

Given the aforementioned background, CPU performance prediction has become an important technology for CPU design and management \cite{hennessy2019new}. This technology can predict the CPU performance based on various hardware characteristics. For CPU manufacturers, effective CPU performance prediction methods enable hardware designers to efficiently obtain prototype parameters, facilitating the design of high-performance CPUs or those tailored to specific application scenarios, thereby maximizing production efficiency and economic benefits. For purchasers and consumers, these methods allow for the quick and easy selection of CPUs that meet their specific requirements, enhancing hardware procurement efficiency. In summary, CPU performance prediction possesses substantial theoretical research value and practical application significance, representing one of the most valuable research areas.

There exist two main challenges in the field of CPU performance prediction. First, there is a lack of a representative, significant, and comprehensive dataset with large-scale data and meaningful evaluation metrics. Currently, most of the CPU benchmark performance data originates from internal testing by manufacturers or professional evaluations by third parties. This data may not be publicly released, or the formats of released data vary significantly, making the collection and standardization of data for prediction exceedingly difficult. Without a large amount of actual collected data with unified data format and evaluation standards, existing methods cannot effectively predict the performance of CPUs with different configurations. Second, the traditional hardware performance evaluation approaches often suffer from the significant limitations. Concretely, the hardware simulation-based approaches usually select some parameter configurations related to CPU performance and calculate performance scores by simulation. Due to the high complexity, this type of approaches usually consumes substantial simulation resources, making them unsuitable for rapid computation and large-scale data analysis \cite{joseph2006predictive}. Moreover, the machine learning-based approaches such as linear regression (LR) \cite{tranmer2008multiple} and support vector machines (SVM) \cite{hearst1998support} cannot guarantee the prediction accuracy, especially when there exist many influencing indicators.

According to the above analysis, we make contributions from two aspects to promote the development of the CPU performance prediction field. First, we organize a novel dataset for further research. Concretely, we collect the historical CPU benchmark data of the 4th Generation Intel\textsuperscript{\textregistered} Xeon\textsuperscript{\textregistered} Scalable Processors, including data samples with 83-dimensional the hardware characteristics and the 1-dimensional corresponding performance prediction scores under different benchmark suites. Following the data cleaning, data standardization and feature engineering processes, we can generate standard data instances in the dataset. As a result, we organize a novel CPU performance prediction dataset called PerfCastDB, including totally $13048$ instances. Each instance contains $35$ hardware characteristics, and $1$ ground truth prediction scores under $6$ testing suites. Note that as the design models and historical performance data of Intel CPUs are still being accumulated, the data scale of this database will continue to expand. Second, we propose a novel deep learning based approach as the baseline on the organized benchmark. It divides the indicators into different groups according to their physical properties. The inter-group and intra-group correlations are comprehensively modeled to focus on the indicators with significant influence to the prediction results. For better understanding of the organized data, we provide a sub-benchmark sample at the link \href{https://github.com/xiaoman-liu/NCPP/blob/main/data/raw/SPR/train_data.csv}{Intel Sapphire Rapids sample}.

In general, the contributions of this article can be summarized as:
\begin{itemize}
	\item We propose a comprehensive benchmark PerfCastDB, which suits for the CPU performance prediction task. We have systematically collected, organized and standardized large-scale data on CPU physical characteristics and prediction performances under different benchmark suites. The proposed PerfCastDB dataset can provide effective data support for the subsequent research on CPU performance prediction tasks.
	\item We propose a novel NCPP as the baseline on the proposed PerfCastDB dataset. NCPP divides several impact hardware characteristics into different groups, and the intra-group and inter-group interactions are effectively modeled and utilized for instance-wise feature learning. NCPP presents a systematic analysis and dynamic adjustments on the affecting hardware characteristics to improve the prediciton effectiveness.
	\item We compare the proposed NCPP with several traditional approaches. The experimental results show the superiority of NCPP in terms of CPU performance prediction, which evaluates the effectivenss of NCPP.
\end{itemize}

The rest of this paper is organized as follows. In Sec.2, we introduce the related work. Sec.3 presents the organization details of PerfCastDB dataset. Sec.4 shows the design of the NCPP network. In Sec.5, we conduct comprehensive experiments to demonstrate the effectiveness of NCPP. Finally, we conclude this article in Sec.6.

\section{Related Works}

Numerous studies have focused on understanding and predicting computer system performance. Early work \cite{ein1987attributes, Rafael1996} utilized statistical analysis to explore the relationship between CPU performance and its physical characteristics. Benjamin et al. \cite{lee2007illustrative} and Nussbaum et al. \cite{nussbaum2001modeling} employed statistical and sampling methods to analyze computer performance, thus avoiding the time-consuming and resource-intensive process of precise simulations. Subsequently, research shifted towards using machine learning methods for CPU performance prediction. Hamerly et al. \cite{hamerly2006using} proposed the SimPoint tool, which employs data clustering algorithms to automatically identify repetitive patterns during program execution. Lin et al. \cite{lin2013application} significantly improved the accuracy of CPU performance prediction by combining LS-SVR with the PSO method and also explored the application of GRNN and RBNN methods. Malakar et al. \cite{malakar2018benchmarking} evaluated $11$ machine learning methods across $4$ scientific applications and platforms, demonstrating promising results for bagging, boosting, and deep neural network methods. Additionally, Mankodi et al. \cite{mankodi2020evaluating} and Tousi et al. \cite{tousi2022comparative} studied the effectiveness of various traditional machine learning models in benchmark program performance prediction, finding that tree-based models achieved better accuracy. However, these studies did not evaluate the methods on a comprehensive and diverse CPU performance dataset and did not overcome the performance limitations of traditional models.

The explosive growth of deep learning in text \cite{vaswani2017attention}, speech \cite{hinton2012deep}, and image recognition \cite{he2016deep} has introduced new avenues for performance prediction \cite{dean2018new}. Dibyendu et al. \cite{das2018specnet} introduced the deep neural network SpecNet, achieving high accuracy in prediction. Yu Wang et al. \cite{wang2019predicting} explored the performance relationship between different benchmark tests using deep neural network models, showing significant precision improvement compared to traditional linear models. Michael et al. \cite{siek2023benchmarking} constructed a long short-term memory (LSTM) model to predict the performance of CPUs and GPUs. Cengiz et al. \cite{cengiz2023predicting} demonstrated the potential of deep learning models in predicting SPEC CPU 2017 benchmark performance, with convolutional neural networks achieving a higher $R^2$ value. Although these models have an advantage in accuracy, they lack high explainability.

Traditional machine learning methods are easy to implement and interpret but require extensive manual feature engineering and struggle to capture complex data relationships. Traditional machine learning methods such as LR, lasso \cite{tibshirani1996regression}, ridge \cite{hoerl1970ridge}, elastic net (EN) \cite{zou2005regularization}, SVM, and xgboost (XGB) \cite{chen2016xgboost} are straightforward to implement and interpret. However, they require extensive manual feature engineering and struggle to capture complex data relationships. In contrast, end-to-end representation learning methods such as LSTM \cite{hochreiter1997long}, gated recurrent unit (GRU) \cite{chung2014empirical}, and convolutional neural networks (CNN) \cite{gu2018recent} can automatically learn relationships between features and model complex non-linear relationships. However, they are prone to overfitting and require large amounts of data and computational resources. To avoid the need for extensive manual and time-consuming feature engineering, we design an end-to-end model to learn the relationship between input hardware characteristics and output.

Building on the foundations of traditional machine learning and end-to-end representation learning methods, our work introduces a significant advancement in the field. Our model, based on deep learning techniques, can automatically learn complex feature representations compared to traditional machine learning methods. Additionally, we introduce the attention mechanism to the CPU performance prediction field for the first time. This mechanism captures key features affecting CPU performance, thereby enhancing model explainability. To demonstrate the effectiveness of our model, we conduct extensive experiments, comparing it against seven existing machine learning methods that have proven effective in performance prediction in previous research. In most cases, our model demonstrates superior performance.

\section{Dataset Organization}
In this section, we introduce the organization process of PerfCastDB, including data collection, data preprocessing, benchmark suites, and feature engineering.

\subsection{Data Collection}
The data is collected from Sapphire Rapids (SPR) \cite{Products57}, the 4th Generation Intel\textsuperscript{\textregistered} Xeon\textsuperscript{\textregistered} Scalable Processors based on Intel 7 technology. Data collection commenced on September 27, 2022, and continued until October 27, 2023, covering a variety of stock-keeping units (SKU) within the SPR product line during this period. Our model is exposed to a representative variety of processors, providing a robust foundation for analyzing performance across different configurations and capabilities.

These suites focus on evaluating different factors affecting CPU performance, leading to significant differences in the corresponding CPU performance scores under the same hardware characteristic standards. During the data collection process, we fully consider the advantages of different suites, and select four mainstream benchmark suits for dataset organization. As a result, we organize $6$ sub-suites under $4$ benchmarks suites, two of which contain two sub-suites respectively. The benchmark suites and their functions are detailed in Table \ref{tab:sumary}.

\begin{table}[htbp]
	\centering
	\caption{Overview of the testing suites utilized during the data collection process}
	\begin{tabular}{@{}ccp{6.8cm}@{}}
		\toprule
		\textbf{Suite Name}   &\textbf{Bnechmarks}    & \textbf{Description}                                                                                  \\ \midrule
		SPECrate2017\_int\_base          & 11         & it measures the integer computation capabilities of a CPU                                                 \\ 
		\addlinespace
		SPECrate2017\_fp\_base           & 14         & it measures the floating-point computation capabilities of a CPU                                          \\ 
		\addlinespace
		MLC Latency       & 9             & it measures the latency for CPU to access data from cache or memory                     \\ 
		\addlinespace
		MLC Bandwidth        & 9           & it measures the bandwidth for CPU to access data from cache or memory                     \\ 
		\addlinespace
		Stream                    & 4                & it measures sustainable memory bandwidth and computation rate for simple vector kernels \\ 
		\addlinespace
		HPCG                         & 1              & it measures the computational performance of solving sparse matrix equations                              \\ \bottomrule
	\end{tabular}
	\label{tab:sumary}
\end{table}

Specifically, SPEC CPU2017 \cite{SPECCPU90} consists of two main metrics: SPECrate and SPECspeed. Here, we take the SPECrate as a suite, which can be divided into "SPECrate2017\_int\_base" and "SPECrate2017\_fp\_base" suites. In a multi-socket system, local memory latencies and cross-socket memory latencies vary significantly \cite{IntelMe18}. Memory Latency Checker (MLC), developed by Intel, is used to measure these local and cross-socket latencies and bandwidth from a specific set of cores to caches or memory. We employ the "idle\_latency" argument to test the latency from initiating a memory request to receiving a response in an idle state, which includes $9$ different levels of latency. For bandwidth, we collect $9$ types of bandwidth data, including the maximum bandwidth of L3, as well as the bandwidth within and across sockets at different read/write rates. Stream \cite{STREAMBe96} uses four metrics for analyzing bandwidth: "Copy", "Scale", "Sum", and "Triad". These metrics collectively provide insights into memory efficiency across various computing scenarios. High Performance Conjugate Gradients (HPCG) \cite{dongarra2016new} is a commonly used benchmark test program for evaluating the performance and efficiency of high-performance computing systems.

\subsection{Data Processing}
\subsubsection{Data Exploration} 

The sub-benchmarks listed in Table.\ref{tab:sumary} share the same hardware characteristics within the same suite. Here we illustrate the numerical characteristics and labels of suite "SPECrate2017\_int\_base" for better understanding.

\begin{figure*}[!t]
	\centering
	\includegraphics[width=\textwidth]{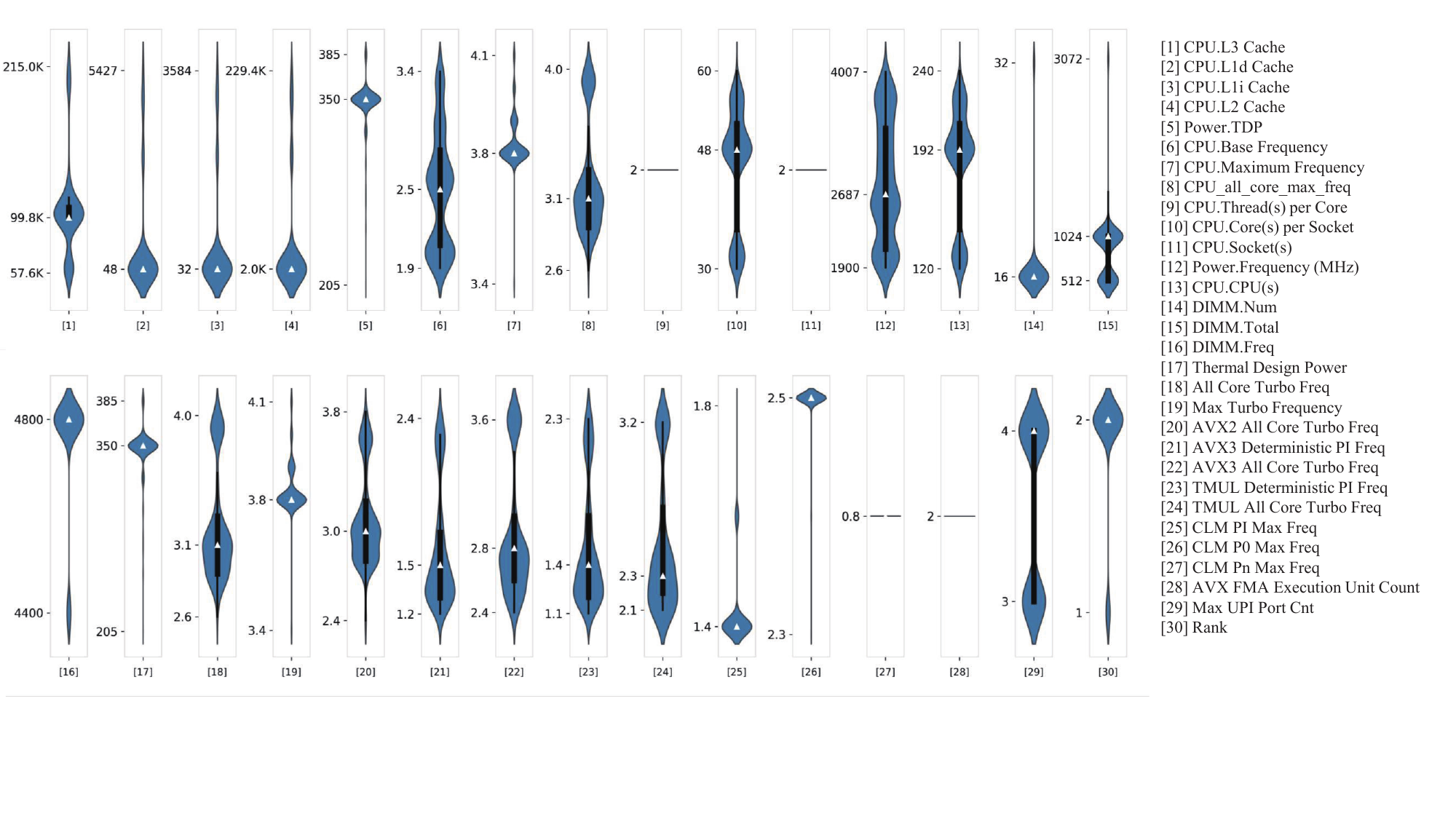}
	\caption{Multivariate Violin Plot Displaying the Distribution of Various CPU Characteristics: From Cache Size to Frequency, as well as Multicore Processing Capabilities and Memory Configuration SPECrate2017\_int\_base}
	\label{fig:violin}
\end{figure*}

Figure.\ref{fig:violin} presents an overview of numerical characteristics, including the minimum, median, and maximum values for thorough analysis. Each subplot illustrates the distribution of a particular feature, with the width indicating the data point frequency across different values. This visual representation offers valuable insights into the features prior to data preprocessing and modeling. For example, features like "CPU.Core(s) per Socket" and "CPU.Base Frequency" exhibit wide distributions, indicating a broad range across different CPU models and suggesting their potential impact on performance. Conversely, "Thermal Design Power" shows a more focused distribution, particularly wider at the value of $350$, suggesting a concentration of values around this figure. Additionally, the visualization allows us to spot hardware characteristics with similar distributions. For instance, the distributions of "CPU.Core(s) per Socket" and "CPU.CPU(s)" are remarkably similar, hinting at a possible correlation between these two characteristics. Indeed, the CPU count can be derived from the formula: $ CPU.CPU(s) = number \ of \ sockets \ * \ CPU.Core(s) \ per \ Socket$. This information can help identify redundant or highly correlated features during the feature selection and engineering process.

\begin{figure}[!t]
	\centering
	\includegraphics[width=0.8\textwidth]{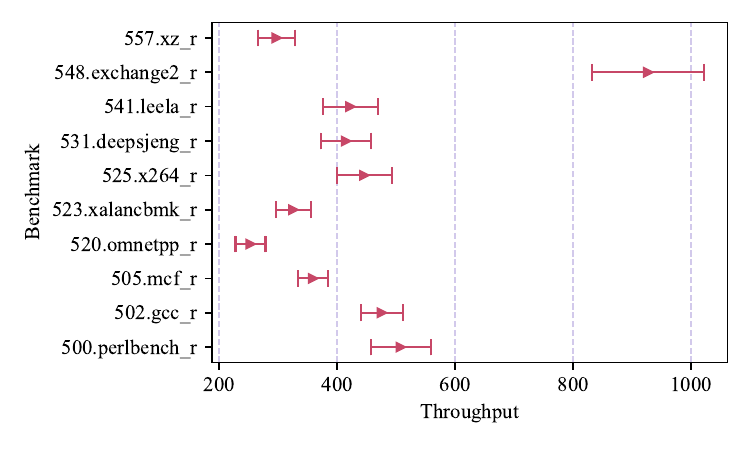}
	\caption{Horizontal Error Bar Chart of Benchmark Performance Measuring Throughput in SPECrate2017\_int\_base}
	\label{fig:errorbar}
\end{figure}

Figure.\ref{fig:errorbar} presents a horizontal error bar graph illustrating the performance throughput of different benchmarks under the "SPECrate2017\_int\_base". The central points represent the mean throughput, while the length of the bars indicates the standard deviation of the throughput. As shown in this figure, performance varies greatly across the different benchmarks, indicating the diversity in computational demands and system utilization in each benchmark. Some benchmarks, like "525.x264\_r", have a large standard deviation, indicating they are more sensitive to changes in hardware characteristics. The application area of the benchmarks can be found in the link \cite{SPECCPU90}.

After collecting the original data, we perform the data processing for data standardization. There exist two main stages: outlier clean and multi-output conversion. Though these steps we obtain a clean and high quality dataset (PerfCastDB) for model training.


\textbf{Outlier Cleaning} We employ a z-score based filtering method to identify and remove outliers of the original data instances. The calculation process of z-score is shown in the following equation: 

\begin{center}
	\begin{equation}
		z = \frac{x - \mu}{\sigma}
		\label{zscore}
	\end{equation}
\end{center}
where $x$ represents the raw scores, $\mu$ is the mean of the scores, and $\sigma$ is the standard deviation of the scores. We set z-score threshold to 3, any data points with a z-score exceeding this threshold ($|z| > 3$) are removed from the benchmark dataset. This method effectively filters out outliers that could distort the analysis, ensuring the quality and accuracy of the dataset.

\textbf{Multi-output Conversion} Our dataset includes six types of benchmark performance data, each with a varying number of benchmarks. For example, in "SPECrate 2017\_int\_base" includes 11 benchmarks, each producing a distinct run record with its own set of features, resulting in 11 separate run records. Despite being tested in a suite run, all benchmarks share the same hardware configuration.

To streamline the data, we consolidate these individual test records into a single record with a 11-dimensional vector as the output. This consolidation simplifies the training process significantly. By transforming the data into a multi-output format, the model only needs to process one-eleventh of the original data, thereby reducing redundant feature training. This optimization leads to faster training times and potentially lower computational costs. Additionally, multi-output learning captures complex relationships between labels, improves predictive performance, and enhances model interpretability.

\subsection{Feature Engineering}\label{sec:features}

\subsubsection{Feature Division}
By leveraging data from these benchmark suites, we can construct a comprehensive profile of Intel SPR performance, encompassing both computational and memory-related aspects. To better understand the hardware characteristics within the dataset and enhance the model learning efficiency, we categorize these characteristics into four main groups. We will briefly discuss these groups below. For detailed data format, please refer to our open-source code \cite{NCPPdata1:online}.

\textbf{Memory Group} Our study includes a set of memory-related characteristics to assist in predicting CPU performance. For a detailed description of the memory related characteristics, please refer to the documentation available on GitHub \cite{NCPPdata1:online}. These features include total memory capacity, the number of dual in-line memory modules (DIMM) used, memory frequency, "DIMM.PartNo", and DIMM ranks. These features are crucial for understanding the memory performance of the system, which can significantly impact overall CPU performance.

\textbf{Workload Group} The second set of features (GitHub \cite{NCPPdata1:online}) relates to workload characteristics, which are critical for understanding how different tasks may impact CPU performance. These features include the benchmark tool name, its version, preset configurations, and specific test names within the benchmark suite.

\textbf{CPU Group} The CPU information (GitHub \cite{NCPPdata1:online}) is detailed through an array of attributes that describe the performance characteristics and architectural specifics of the processors. These include base and maximum frequencies, core and thread counts, and cache sizes across different levels. This comprehensive set of characteristics is critical for understanding and modeling CPU behavior and performance under different computational workloads.

\textbf{Other Group} By grouping theoretical thermal design power (TDP) with other non-performance-related characteristics (such as CPU model and family), we can analyze these characteristics from the perspective of system design and configuration, and understand their impact on overall performance. These characteristics help us understand the CPU architecture and theoretical TDP, which affects power consumption and cooling requirements. While TDP does not directly reflect computational capability, it is crucial for power supply and cooling system design. This group includes the CPU specific model, stepping, and family from SPR, as well as the microcode version, which addresses firmware-level errors and ensures optimal functionality. The entire list is available on GitHub \cite{NCPPdata1:online}.

\subsubsection{Feature optimization}\label{sec:FE}
In this section, we take several key steps to optimize the dataset. These steps include trimming features, expanding features, and normalizing and tokenizing features.

\textbf{Trim Features} First, to reduce the risk of model overfitting and improve generalizability, we identify and remove duplicated hardware characteristics originating from multiple sources. At the same time, we eliminate characteristics that do not hold physical significance. These primarily include string characteristics used to distinguish between different test cases, as well as characteristics like "DIMM.PartNo" and "META.metadata.cscope.qdf0," which serve merely as codenames for memory and CPU.

\textbf{Feature Expansion} Additionally, we leverage "DIMM.PartNo" to enrich our memory-related characteristics by consulting the official websites of leading DIMM manufacturers like Samsung \cite{ModuleDR99}, Hynix \cite{Technica69}, and Micron \cite{SearchRe73}. We add detailed specifications on memory components, including DIMM generation, density, organization, rank, and CAS Latency (CL). These steps ensure that the feature set used for analysis is both streamlined and rich in relevant details.

\textbf{Normalization and Tokenization} After optimizing the features set, we tokenize categorical features, transforming textual data into numerical tokens to enable more effective deep learning training. Additionally, we normalize numerical features to a consistent scale, significantly improving the efficiency of model training and the accuracy of predictions.

\section{Method}

\begin{figure}[!t]
	\centering
	\includegraphics[width=1.1\textwidth]{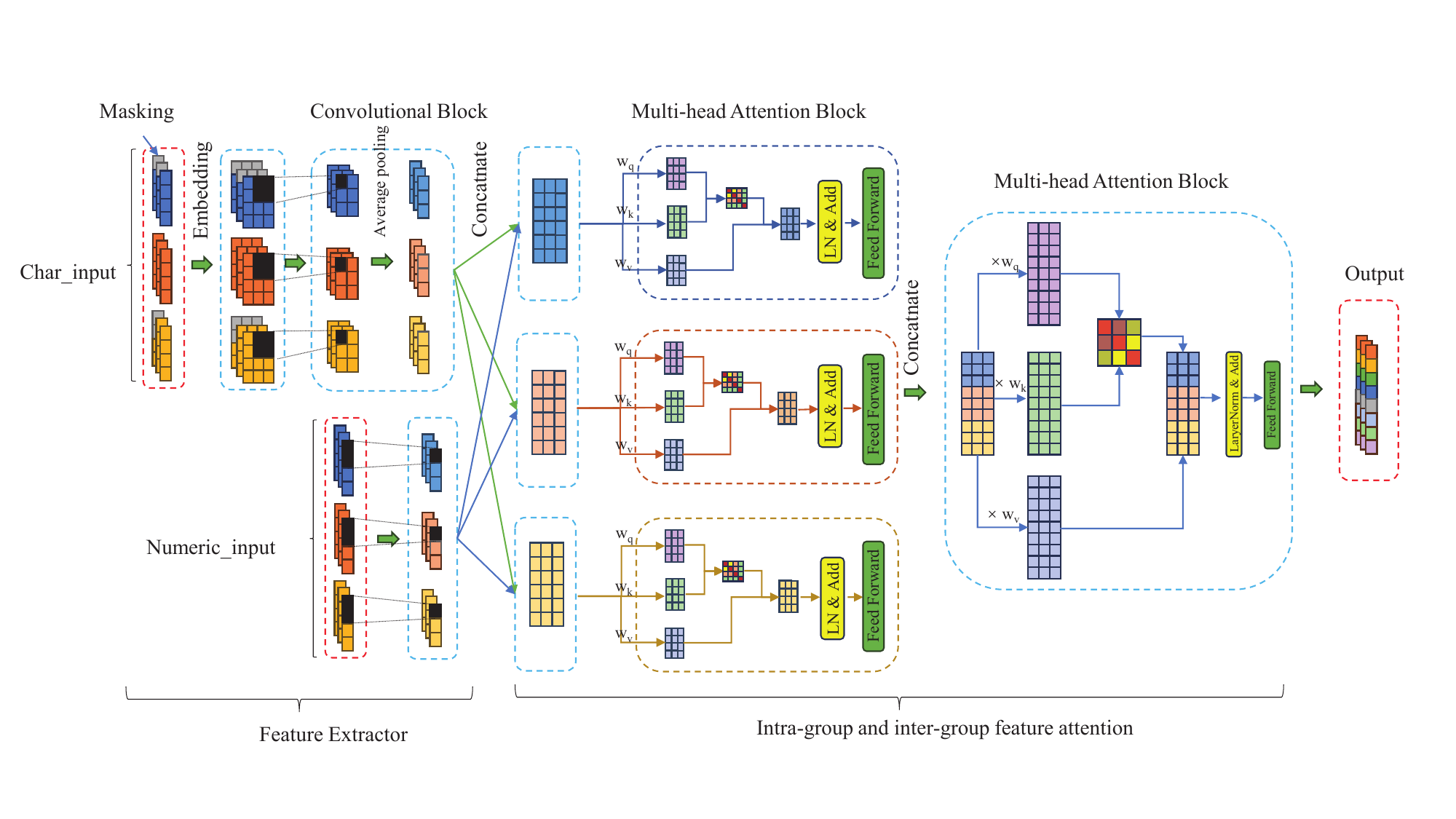}
	\caption{NCPP Architecture: Feature Extraction from Character and Numerical Inputs Using a Deep Learning Model with Convolution and Multiple Attention Mechanisms}
	\label{fig:model}
\end{figure}

Based on the proposed PerfCastDB dataset, we design a novel NCPP as a baseline approach for further research. As shown in Figure. \ref{fig:model}, the architecture of NCPP contains three main components: the feature division module, the intra-group attention module, and the inter-group attention module. We detail these parts in the following paragraphs.

\subsection{Feature Division Module}

\textbf{Characteristic Division}: In this module, we first divide the input characteristics into two main categories: numerical characteristics and the character characteristics. The numerical characteristics include core count, base frequency, and cache size, while the character characteristics encompass the "DIMM.PartNo", benchmark tool name, and benchmark suite name. These characteristics are processed separately before being combined for further analysis. The character characteristics process hardware configuration-related attributes. After tokenizing the character characteristics, the length of the tokens varies. To standardize these lengths across all values, we append zeros to the tokens until they match the maximum token length for each characteristic. To ensure these padded values do not influence the training process, we introduce a mask layer prior to inputting the data into the model. This mask layer effectively ignores the padded zeros during training and is strategically positioned before the embedding layer. The embedding layer functions as a trainable lookup table, assigning each category a fixed-size vector representation. The vocabulary size is set to 100, and we have determined the dimension of the resulting mapped vector to be 4, optimizing the model efficiency in handling character data.

\textbf{Feature Extraction}: Then, we design the convolutional blocks for feature extraction of the two types of characteristics. Specifically, the numerical and character characteristics are processed through separate convolutional blocks, serving as primary feature extractors in the model. These blocks have the capability to skip one or more layers and selectively extract information from various layers through shortcut connections. Each block comprises two sequential layers: a convolutional layer followed by batch normalization and the ReLU activation function. The ReLU activation function introduces non-linearity into the model. The convolutional layers employ a kernel size of 64 x 1. The formula for the shortcut connection can be expressed by:

\begin{equation}
	\begin{aligned}
	{Output} & = \mathcal{F}(x, {W_i}) + x
\end{aligned}
	\label{residual}
\end{equation}

where $x$ represents the input to the block, $F$ denotes the convolutional sequence, $W_i$ signifies the weights of the $i$th convolutional layer, and the addition operation with $x$ integrates the input with the output of the residual function. This mechanism facilitates the creation of deeper architectures by offering an alternative path for gradient flow during backpropagation, thereby aiding in the efficient training of the model.

\textbf{Feature Division}: The extracted features from both character and numerical characteristics are divided into multiple groups (Char Group, CPU Group, Other Group, Memory Group). This division allows the model to handle different types of features separately, enhancing its ability to capture diverse patterns in the data.

\subsection{Intra-group Attention Module}

After generating the feature groups, we employ parameter-independent intra-group attention module for each feature group. The intra-group attention module is designed based on the multi-head self-attention mechanism. Within the attention module, the interactions and the relationships between the intra-group features can be effectively modeled. The output from all attention heads are concatenated to form a comprehensive representation of each feature group.

We take the feature of a instance X as the attention input, the self attention mechanism can be expressed by: 

\begin{equation}
	\begin{aligned}
		Q^{(l)} &= X^{(l)}W_Q^{(l)} \\
		K^{(l)} &= X^{(l)}W_K^{(l)} \\
		V^{(l)} &= X^{(l)}W_V^{(l)} \\
	\text{Intra-Self-Attention}(Q^{(l)}, K^{(l)}, V^{(l)}) &= \text{softmax}\left(\frac{Q^{(l)}(K^{(l)})^T}{\sqrt{d_k}}\right)V^{(l)}
\end{aligned}
	\label{eq:attention}
\end{equation}
where $W_Q^{(l)}$, $W_K^{(l)}$, and $W_V^{(l)}$ are trainable weight matrices for layer $l$, $d_k$ denotes the embedding dimension. Then outputs of the self attention heads are concatenated and linearly transformed:

\begin{equation}
	\begin{aligned}
	\text{Intra-MultiHead}^{(l)}(Q^{(l)}, K^{(l)}, V^{(l)}) = \text{Concat}(\text{head}_1^{(l)}, \ldots, \text{head}_h^{(l)})W_O^{(l)}
\end{aligned}
	\end{equation}
	where
	\begin{equation}
	\text{head}_i^{(l)} = \text{Intra-Self-Attention}(Q_i^{(l)}, K_i^{(l)}, V_i^{(l)})
	\end{equation}
	and $W_O^{(l)}$ is the linear transformation matrix after concatenation for layer $l$.

Then this block is followed by a residual connection and layer normalization. After that the output is passed through a feed-forward network (FFN), including a dense layer, layer normalization and residual connection. The presentation of FFN is as follows:

\begin{equation}
	\begin{aligned}
	F F N & = \operatorname{ReLU}\left(W_1 X^{(l)}+b_1\right)
\end{aligned}
	\label{eq:ffn}
\end{equation}
where $W_1, b_1$ are the weights and biases of the fully connected layer in FFN. Then the normalization layer is introduced as:

\begin{equation}
	\begin{aligned}
		 LN_1 &= \text{LayerNorm}(X^{(l)} + \text{Intra-MultiHead}) \\
		 LN_2 &= \text{LayerNorm}(FFN(LN_1) + LN_1)
	\end{aligned}
	\label{eq:ffn_bn}
	\end{equation}

\textbf{Inter-group Attention Module}
The concatenated outputs from the intra-group attention modules are further processed through an inter-group attention module. Similar to the intra-group module, this module uses multiple attention heads to perform self-attention on the combined feature groups, followed by "Add \& Norm" operations and "Feed Forward" layers in Figure.\ref{fig:model}. The final outputs from these attention heads are concatenated to form a unified feature representation.

\begin{equation}
	\begin{aligned}
	\text{Inter-MultiHead}^{(l)}(Q^{(l)}, K^{(l)}, V^{(l)}) = \text{Concat}(\text{head}_1^{(l)}, \ldots, \text{head}_h^{(l)})W_O^{(l)}
\end{aligned}
	\end{equation}
	where
	\begin{equation}
	\text{head}_i^{(l)} = \text{Inter-Self-Attention}(Q_i^{(l)}, K_i^{(l)}, V_i^{(l)})
	\end{equation}
	and $W_O^{(l)}$ is the linear transformation matrix applied after concatenation for layer $l$. The structure of "Inter-Self-Attention" is similar to that described in Equation \ref{eq:attention}.

This architectural arrangement offers several potential benefits. It allows the model to autonomously learn the characteristics of each feature group, akin to partitioning the data into separate subspaces, thereby enabling the model to capture distinct feature correlations within these subspaces. Moreover, multi-head attention allows the model to focus on different parts of the input sequence simultaneously, enhancing its ability to capture complex patterns and dependencies. Furthermore, within the multi-attention blocks, the model not only captures the importance of features within a group but also investigates the inter-group relationships among features. These inter-group interactions enhance the model comprehension of intricate feature dependencies, contributing to improved performance predictions.

Finally, the output of the inter-group attention module is fed into a dense layer. The number of the units in this dense layer is equal to the number of benchmarks. Then the result of this dense layer will go through a linear activation function. This output represents the prediction of CPU performance by the model based on the learned features from the input data in PerfCastDB.


\subsection{Loss Function}
In this study, we utilize the huber loss instead of Mean Absolute Error (MAE) or Mean Squared Error (MSE) due to its robustness to outliers. The formula of huber loss can be expressed by:

\begin{equation}
	\text{huber loss} = \left\{\begin{array}{ll}
		\frac{1}{2} \times\left(y-\hat{y}\right)^{2}             & \text { for }|y-\hat{y}| \leq \delta \\
		\delta \times\left(|y-\hat{y}|-\frac{1}{2} \delta\right) & \text { otherwise }
	\end{array}\right.
	\label{eq:huber}
\end{equation}
where $y$ is the true value, $\hat{y}$ is the predicted value, and $\delta$ is a hyperparameter that determines the threshold for the transition between the MAE and MSE loss functions. For errors smaller than the threshold $\delta$, this loss function behaves like MSE and thus is sensitive to large errors and promoting smoothness in the prediction. For errors larger than $\delta$, it acts like MAE, which is less sensitive to large errors and thereby reduce the impact of outliers. The adoption of the huber loss function is particularly beneficial in our study due to the identified presence of outliers in our dataset. This ensures that our model is not overly impacted by these extreme values, leading to more reliable performance predictions for CPUs.

\begin{table*}[!htbp]
	\centering
	\caption{The data distribution for the training, validation, and testing sets on the PerfCastDB dataset under different suites.}
	\begin{tabular}{llll}
		\toprule
		\textbf{Suite Name}              & \textbf{Trainset} & \textbf{Validationset} & \textbf{Testset} \\ \midrule
		SPECrate2017 Integer base        & 816                    & 204                         & 254                   \\
		SPECrate2017 FP base             & 756                    & 189                         & 236                   \\
		Memory Latency Checker Latency   & 140                    & 35                          & 42                    \\
		Memory Latency Checker Bandwidth & 922                    & 230                         & 294                   \\
		Stream                           & 2858                   & 714                         & 897                   \\
		HPCG                             & 2815                   & 704                         & 886                   \\\bottomrule
	\end{tabular}
	\label{tab:dataset}
\end{table*}

\section{Experiments}

\subsection{Data Splitting and Cross-Validation}

As mentioned in the dataset organization subsection, we collect large-scale data under $6$ suites. We then divide the data into three segments: $60\%$ allocated for the training set, $20\%$ for the validation set, and the remaining $20\%$ for the testing set. The specific distribution of sample numbers for each suite is detailed in Table.\ref{tab:dataset}. Considering the different attributes between these suites, we fine-tune the parameters of the NCPP model under each suite for better prediction performance. To ensure effective evaluation, we perform a 5-fold cross-validation. Following the preprocessing phase, we divide the dataset into five equal parts. Each part is used as the validation set in rotation, with the remaining four parts serving as the training set. The final model evaluation is based on the average results from the five validation rounds. This approach helps prevent overfitting and optimizes hyperparameters, thereby enhancing the model's generalization to unseen data.

\begin{table*}[!htbp]
	\centering
	\caption{The data distribution for the training, validation, and testing sets on the PerfCastDB dataset under different suites.}
	\begin{tabular}{llll}
		\toprule
		\textbf{Suite Name}              & \textbf{Trainset} & \textbf{Validationset} & \textbf{Testset} \\ \midrule
		SPECrate2017 Integer base        & 816                    & 204                         & 254                   \\
		SPECrate2017 FP base             & 756                    & 189                         & 236                   \\
		Memory Latency Checker Latency   & 140                    & 35                          & 42                    \\
		Memory Latency Checker Bandwidth & 922                    & 230                         & 294                   \\
		Stream                           & 2858                   & 714                         & 897                   \\
		HPCG                             & 2815                   & 704                         & 886                   \\\bottomrule
	\end{tabular}
	\label{tab:dataset}
\end{table*}

\subsection{Evaluation Metrics}
We utilize the MAE, MSE, and mean absolute percentage error (MAPE) as the evaluation metrics. These metrics can be defined as the following equations:

\begin{equation}
	\begin{aligned}
		\text{MAE} & = \frac{1}{n} \sum_{i=0}^n \left| x_i - \hat{x}_i \right| \\
		\text{MSE} & = \frac{1}{n} \sum_{i=1}^{n} (x_i - \hat{x}_i)^2 \\
		\text{MAPE} & = \frac{1}{n} \sum_{i=1}^{n} \left| \frac{x_i - \hat{x}_i}{x_i} \right|
	\end{aligned}
	\label{combined}
\end{equation}
where $x_i$ represents the ground truth value and $\hat{x}_i$ denotes the model predicted value, with $n$ being the total number of samples. The MAE metric is particularly interpretable as it directly quantifies the average error in the same units as the output variable. In contrast, MSE is more sensitive to outliers due to the squaring of the error terms. MAPE expresses the prediction errors as a percentage of the actual values, facilitating easier interpretation and comparison across different benchmarks and prediction models. In addition to MAE, MSE, and MAPE, we also consider the 95th percentile of Absolute Error (AE), Squared Error (SE), and Absolute Percentage Error (APE) to provide a more comprehensive evaluation.

\subsection{Implementation Details}

Our NCPP is developed based on Keras platform, integrated within TensorFlow $2.10$. All the experiments are conducted on a high-performance computing system. This system is powered by Intel SPR and High Bandwidth Memory (HBM) technology, operating on an $x86_64$ architecture and equipped with $120$ physical cores. These processors are distributed across two sockets, each containing $60$ physical cores, with every core supporting two threads. The processor frequency ranges from a minimum of 800 MHz to a maximum of 3600 MHz and includes a third-level cache of 225 MB.

For optimization, we employ the Adam optimizer, renowned for its effectiveness in optimization tasks. To enhance training stability and prevent the model from overshooting the minimum, we utilize an ExponentialDecay strategy for the learning rate. This approach ensures that the learning rate starts higher and gradually decreases, allowing for larger updates early in training and finer adjustments as training progresses. The initial learning rate is set at 0.01, with a batch size of 64. The model undergoes training over 1000 epochs.

To assess the efficacy of our model, we compare it against several baseline methods, including both machine learning and deep learning techniques. Among the machine learning methods, Lasso is noted for its ability to handle multicollinearity and produce simpler, more interpretable models. Ridge regression also addresses multicollinearity and is less prone to overfitting compared to least squares methods. EN combines the feature selection advantages of Lasso with the multicollinearity handling capabilities of Ridge. SVM are particularly effective in high-dimensional spaces, while XGB can handle missing data and provides feature importance scores. In the realm of deep learning, LSTM networks are capable of learning long-term dependencies through gating mechanisms. GRU simplify the LSTM architecture by using fewer gates and parameters. CNN are efficient for processing spatial data. In the following sections, we present the experimental results comparing the performance of our NCPP model with the aforementioned baseline methods

\subsubsection{Comparison of Prediction Performance}
\begin{figure*}[!t]
	\centering
	\includegraphics[width=1\textwidth]{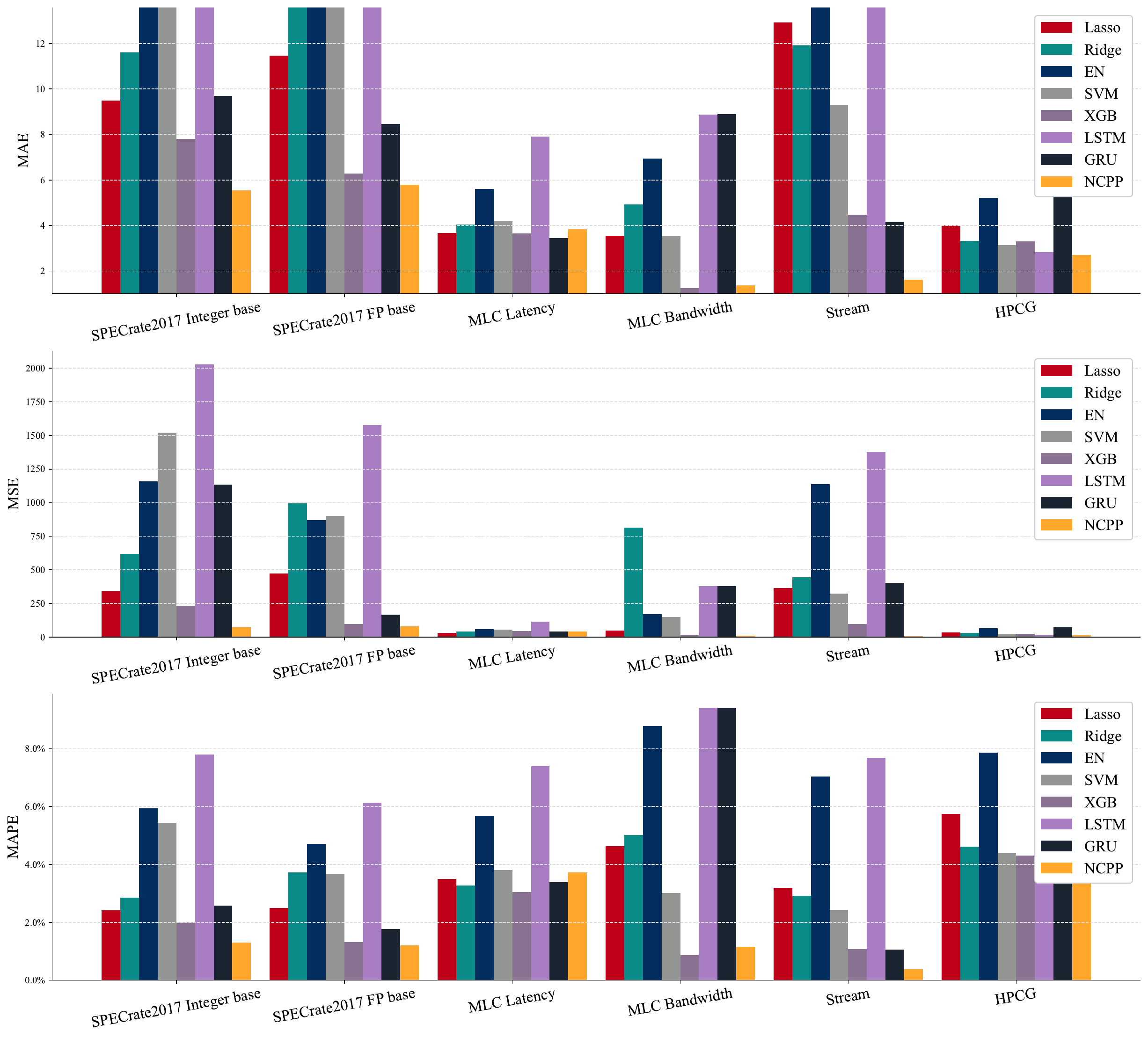}
	\caption{The performance prediction results of NCPP and the comparison apoproaches on the MAE, MAPE, and MSE metrics under different benchmark suites.}
	\label{fig:model_perf}
\end{figure*}

\begin{figure}[!t]
	\centering
	\includegraphics[width=1\textwidth]{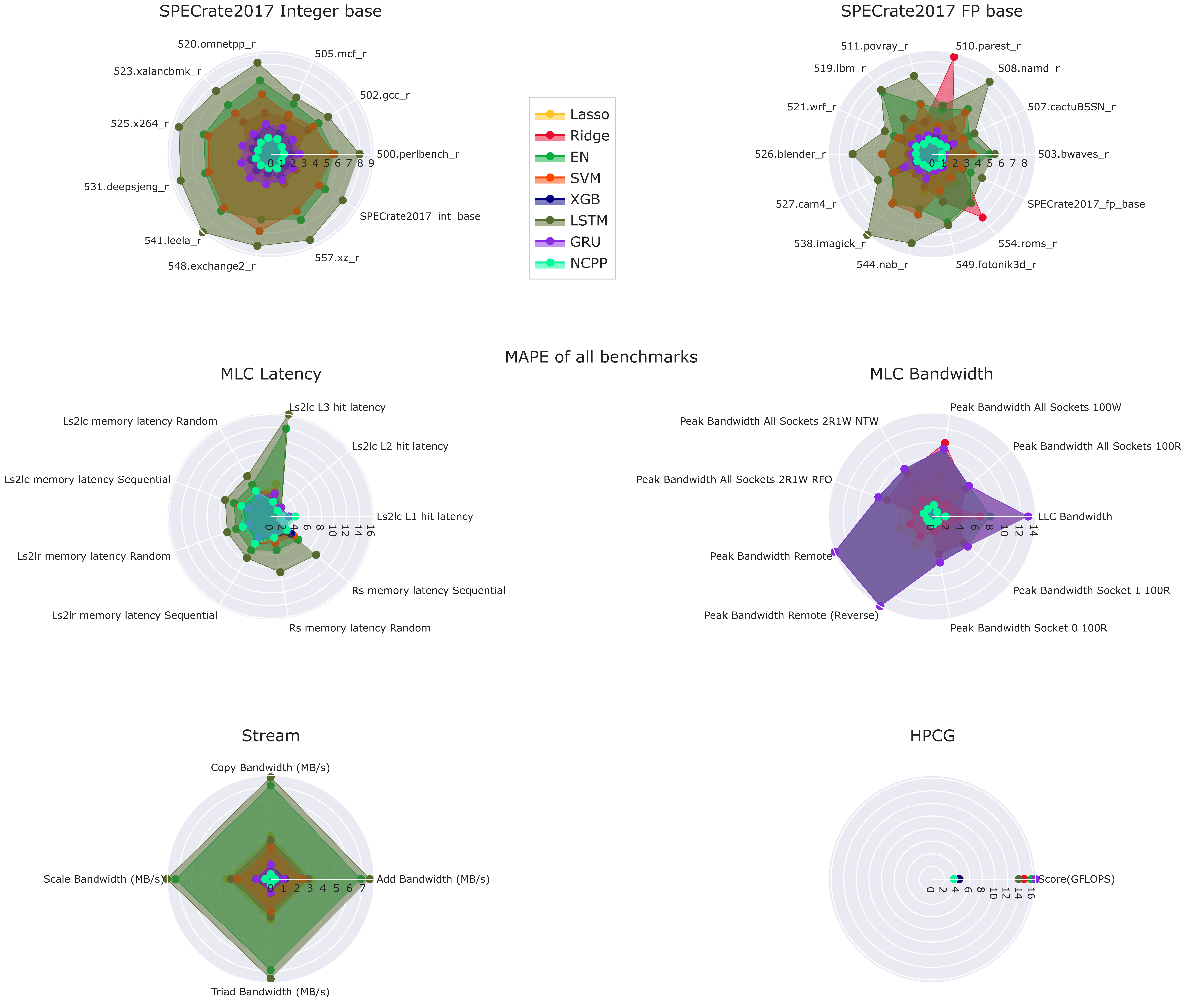}
	\caption{Detailed Performance Evaluation: Radar Chart Comparison of MAPE for NCPP and Conventional Machine Learning Models in SPECrate2017 Integer and Floating Point Benchmarks, MLC Latency and Bandwidth, Stream, and HPCG}
	\label{fig:model_perf_bench}
\end{figure}

To validate the effectiveness of NCPP, we compare it with some representative approaches on each suite. The corresponding results are shown in Figure.\ref{fig:model_perf} and Figure.\ref{fig:model_perf_bench}. As a result, NCPP achieves the best evaluation results on most suites, which directly reflects the superiority of it. Moreover, we discuss the experimental results in detail as the following paragraphs show.

Figure.\ref{fig:model_perf} illustrates the overall prediction performances of NCPP and the comparison approaches. Due to the significantly larger magnitude of bandwidth values measured by MLC compared to other benchmark suites, we scale the real scores and prediction results of MLC by a certain proportion to display them on the same figure as other benchmark suites. Several key observations can be made: 

\begin{enumerate}
	\item Models based on linear relationships, such as LR, Lasso, Ridge, and EN, exhibit subpar performance across all six benchmark suites in our dataset. This indicates that these models are inadequate for capturing the complex dynamics present in PerfCastDB.
	\item Models that utilize end-to-end representation learning, such as GRU, demonstrate better performance than those based on linear relationships. This suggests that the relationship between the features and the outcomes in our dataset is strongly non-linear, and solely relying on linear relationships for prediction is insufficient.
	\item Our model outperforms other methods on most benchmarks, highlighting its superior capability to learn and represent the intricate relationships between features and results. This achievement underscores the effectiveness of NCPP and its potential as a powerful tool for CPU performance prediction.
\end{enumerate}

Figure.\ref{fig:model_perf_bench} provides a detailed view of the prediction performances across various benchmarks. Each axis on the radar chart corresponds to a different benchmark metric, with the center indicating the optimal minimum MAPE value of zero. A point's proximity to the center reflects a lower MAPE thus indicating superior predictive accuracy. From this analysis, we can draw two primary observations.

\begin{enumerate}
	\item The NCPP demonstrates remarkable stability across a wide range of benchmarks, notably including "$SPECrate2017\_fp\_base$" and "$SPECrate2017\_int\_base$". This stability suggests that our model is adept at handling both floating point and integer benchmarks, making it a versatile tool for selecting CPUs optimized for various types of workloads. This adaptability is crucial for ensuring that performance predictions are relevant and applicable to a broad spectrum of computational tasks.

	\item In the case of the MLC, the NCPP performance is comparable to that achieved by XGB, without showing a significant enhancement. This may be attributed to two main reasons: First, the training set for MLC latency is extremely limited, containing only 140 samples. This scarcity of data restricts the fitting capability of the NCPP. Second, the MLC suite tests latency and throughput in an idle state, where characteristics are relatively simple and cannot fully leverage the advantages of the NCPP.
\end{enumerate}

\subsubsection{Prediction Error Analysis}

\begin{figure}[!t]
	\centering
	\includegraphics[width=\textwidth]{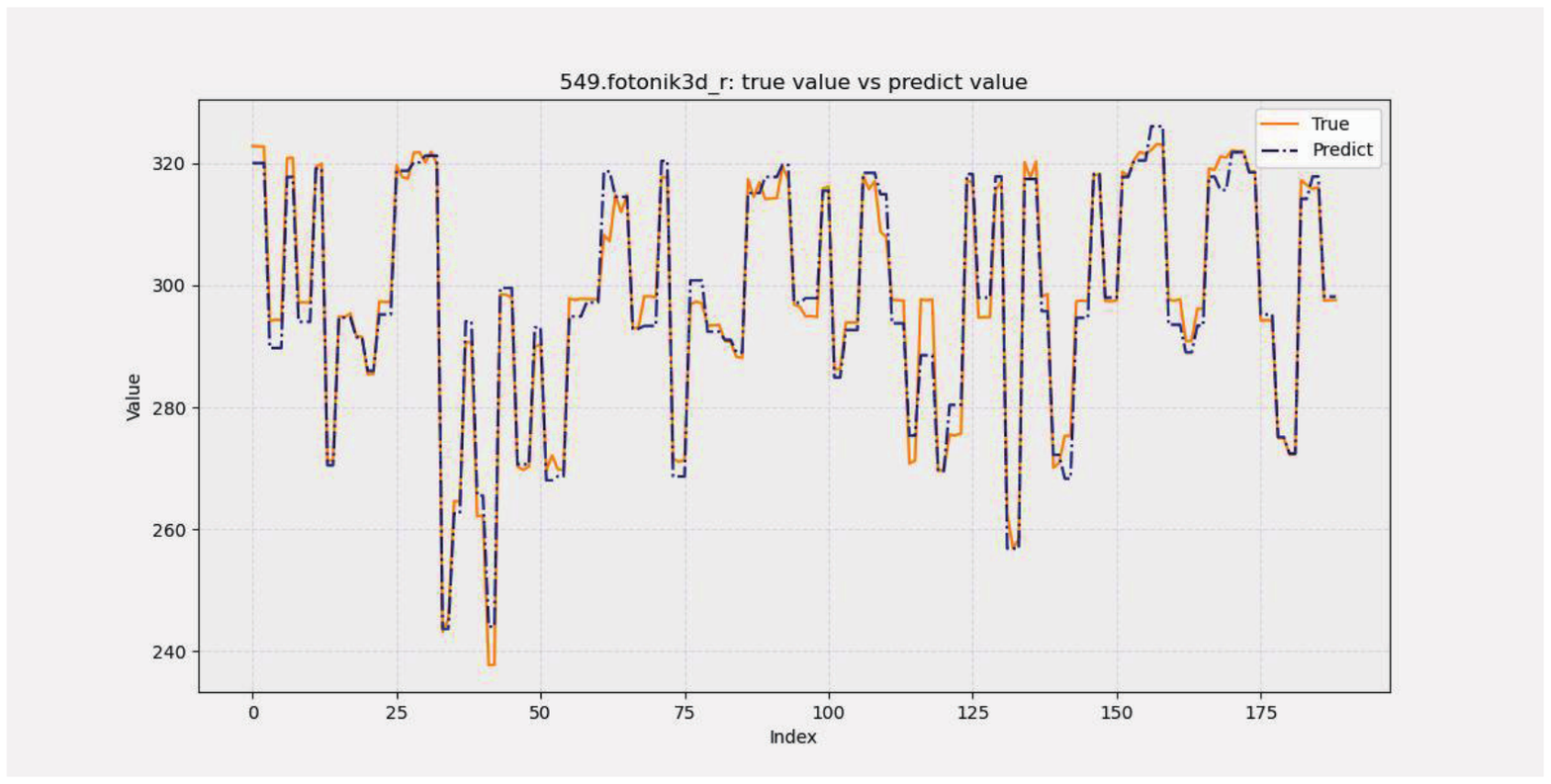}
	\caption{Prediction visualization of $544.nab\_r$ in $SPECrate2017\_fp\_base$}
	\label{fig:errors}
\end{figure}

To intuitively demonstrate the effectiveness of NCPP in CPU performance prediction, we visualize the prediction errors of NCPP on the test set of a specific benchmark suite. Figure.\ref{fig:errors} shows the comparison between the true and predicted values of the NCPP on the test set of "544.nab\_r" in "SPECrate2017\_fp\_base". The x-axis represents the index of the test data, ranging from 0 to 236, while the y-axis displays the true and predicted values. Across the majority of data points, the prediction errors of NCPP is within a reasonable range, achieving a MAPE of less than $4.18\%$.

\begin{table}[!ht]
	\centering
	\caption{Evaluating the Impact of Intra-Group and Various Inter-Group Attention Mechanisms on Model Accuracy: An Ablation Study with MAE, MSE, and MAPE Outcomes}
	\begin{tabular}{cp{3cm}p{3cm}p{3cm}}
		\toprule
		\textbf{Experiments} & \textbf{MAE} & \textbf{MSE} & \textbf{MAPE} \\ \midrule
		(1)                  & 5.79         & 80.58        & 1.20\%        \\
		(2)                  & 7.27         & 144.12       & 1.44\%        \\
		(3)                  & 10.03        & 219.95       & 1.99\%        \\
		(4)                  & 78.30        & 8134.78      & 15.95\%       \\
		(5)                  & 65.64        & 5457.15      & 13.41\%       \\
		(6)                  & 7.90         & 178.06       & 1.55\%        \\
		\bottomrule
	\end{tabular}
	\label{tab:atten_ab}
\end{table}

\subsection{Ablation Study}

In this subsection, we conduct the ablation study to demonstrate the effectiveness of each part in NCPP. Specifically, the main components in NCPP are attention blocks, including the intra-group attentions and inter-group attentions. Based on the baseline NCPP architecture, we make different combinations of intra-group and inter-group attention mechanisms, the corresponding results are shown in Table. \ref{tab:atten_ab}. The network definition under different conditions are explained as follows:

(1) NCPP baseline model: This configuration incorporates both intra-group and inter-group attention mechanisms, serving as the baseline for our comparative analysis.

(2) NCPP without intra-group attention: In this variation, we omit the intra-group attention mechanism while retaining the inter-group attention. This modified model is trained using the same dataset and hyperparameters as the baseline for a fair comparison.

(3) NCPP without "memory group attention": Here, we remove the attention mechanism specific to the memory-relevant group, while keeping both intra-group and other inter-group attentions.

(4) NCPP without "other group attention": In this setup, we exclude the attention mechanism dedicated to the other group mentioned in the above paramgrapghs, maintaining the intra-group and other inter-group attentions.

(5) NCPP without "CPU group attention": This configuration leaves out the CPU group attention mechanism while keeping the intra-group and other inter-group attentions active.

(6) NCPP without "workload group attention": Finally, we remove the attention mechanism for the workload group, leaving the intra-group and other inter-group attentions in place.

In summary, NCPP achieves the best results by involving all the group attention mechanisms, which reflects these attentions have a positive gain effect. Moreover, the microarchitectural and thermal group attention mechanisms in other group appear to be the most critical for maintaining low prediction errors, followed by the CPU and memory group attentions. While the intra-group and workload group attentions also contribute to model performance, their impact is comparatively lesser. This analysis underscores the importance of attention mechanisms in handling complex feature interactions for more accurate CPU performance prediction.

\subsection{Hyperparameters Study}
In this subsection, we conduct the comparison experiments on different settings of important hyperparameters, focusing on the number of attention heads $H$, the number of attention layers $L$, and the weight of huber loss $\delta$ on MAE, MSE and MAPE. We test the number of attention heads $H$ ranging from $1$ to $5$ and the number of attention layers $L$ also from $1$ to $5$, and the value of $\delta$ is chosen from $\{0.1, 1, 10\}$. Table. \ref{tab:hyperparam} presents the corresponding results on "SPECrate2017\_FP\_base" benchmark suite.

\begin{table}[htbp]
	\centering
	\caption{Optimization of Model Hyperparameters for SPECrate2017 FP: Analysis of Layer and Head Configurations, and Huber Loss Impact on MAE, MSE, and MAPE}
	\begin{tabular}{cp{2cm}p{2cm}p{2cm}p{2cm}}
		\toprule
		\textbf{Hyperparameter} & \textbf{Value} & \textbf{MAE}  & \textbf{MSE}   & \textbf{MAPE}   \\ \midrule
		layer=1, head=H         & 1              & 6.29          & 121.52         & 1.29\%          \\
		~                       & \textbf{2}     & \textbf{5.79} & \textbf{80.58} & \textbf{1.20\%} \\
		~                       & 3              & 6.59          & 96.92          & 1.36\%          \\
		~                       & 4              & 7.01          & 118.86         & 1.46\%          \\
		~                       & 5              & 6.58          & 99.39          & 1.36\%          \\ \midrule
		head=2, layer=L         & \textbf{1}     & \textbf{5.79} & \textbf{80.58} & \textbf{1.20\%} \\
		~                       & 2              & 7.74          & 212.57         & 1.64\%          \\
		~                       & 3              & 6.94          & 163.91         & 1.44\%          \\
		~                       & 4              & 7.93          & 137.74         & 1.66\%          \\
		~                       & 5              & 7.05          & 121.02         & 1.46\%          \\ \midrule
		huber loss (delta=D)    & 0.1            & 7.45          & 306.49         & 1.51\%          \\
		~                       & \textbf{1}     & \textbf{5.79} & \textbf{80.58} & \textbf{1.20\%} \\
		~                       & 10             & 7.85          & 180.71         & 1.55\%          \\ \bottomrule
	\end{tabular}
	\label{tab:hyperparam}
\end{table}

From the experimental results, the NCPP model performs best when the number of heads ($H$) is set to 2 and the number of layers ($L$) is set to 1. Specifically, this configuration achieves the lowest MAE and MSE values of 5.79 and 80.58, respectively, with a MAPE of 1.20\%. Increasing the number of heads and layers does not significantly improve the experimental results; instead, it increases the computational complexity of the model. These findings suggest that a simpler model configuration with fewer layers and heads is more effective. Adding more layers or heads tends to result in higher error rates, indicating potential overfitting or increased model variance.

The impact of varying the delta parameter of the huber loss function was also examined. The results show that a delta value of 1 achieves the best performance, with an MAE of 5.79, MSE of 80.58, and MAPE of 1.20\%. This indicates that the model benefits from a balanced approach to handling outliers, as provided by the huberloss with delta set to 1. In contrast, both lower (delta=0.1) and higher (delta=10) values result in significantly worse performance, with higher error rates across all metrics. This highlights the importance of carefully tuning the Huber loss parameter to achieve optimal model performance.

In summary, the hyperparameter optimization results suggest that a simpler model configuration with fewer layers and heads, combined with an appropriately tuned Huber loss function, leads to the best predictive performance. These findings underscore the importance of balancing model complexity and robustness to outliers in achieving accurate and reliable predictions.

\section{Visualization Results}\label{sec12}

The group attention mechanisms are essential parts in NCPP, which can effectively capture and model the impacts and relationships between the features. Here, we respectively visualize the intra-group and inter-group attention relation matrices and the calculated importance matrix of the attention mechanisms.

\begin{figure*}[htbp]
	\centering
	\begin{minipage}[b]{0.45\textwidth}
		\includegraphics[width=\textwidth]{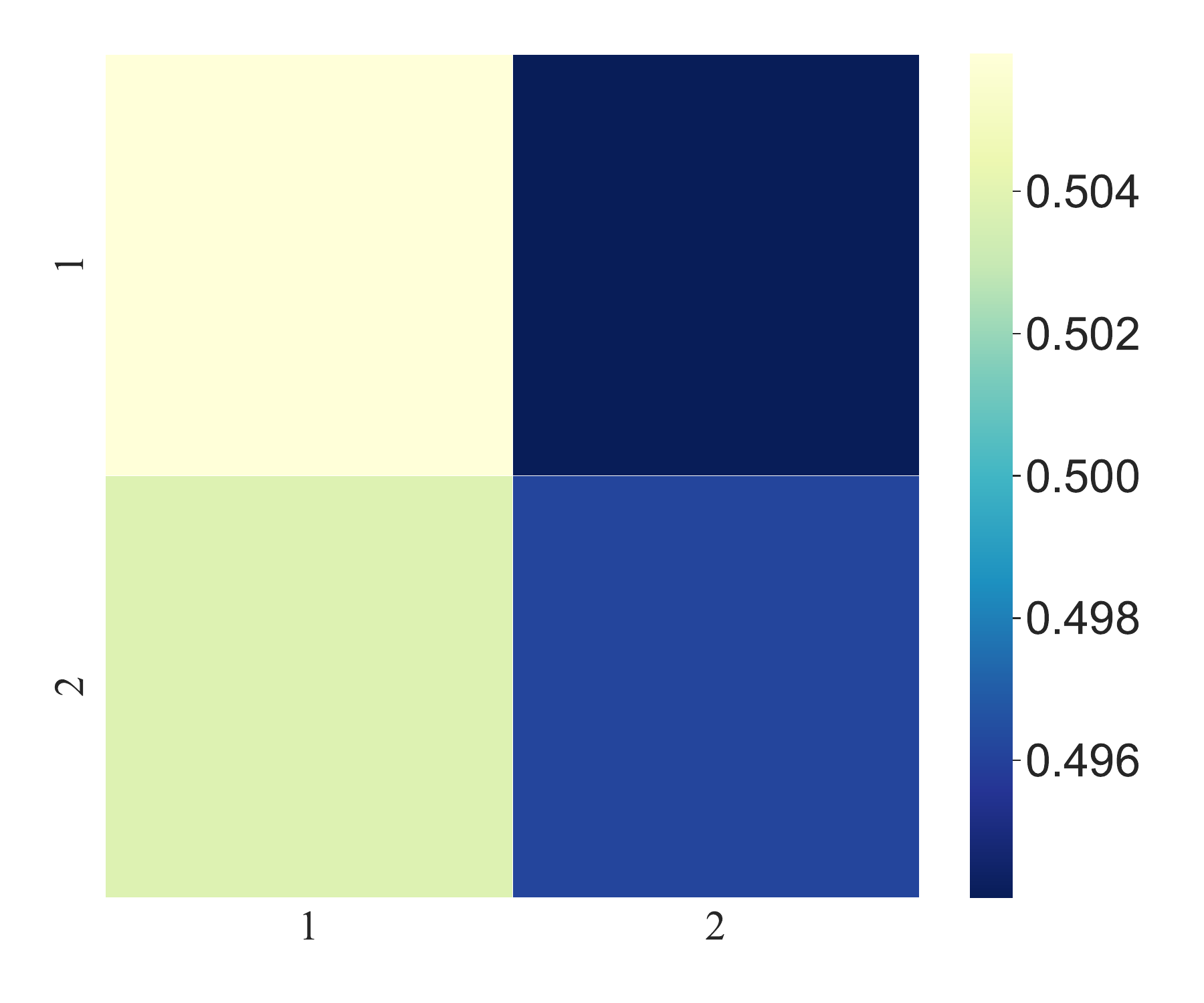} 
		\subcaption{Other Group}
		\label{fig:1_system_matrix}
	\end{minipage}
	\begin{minipage}[b]{0.45\textwidth}
		\includegraphics[width=\textwidth]{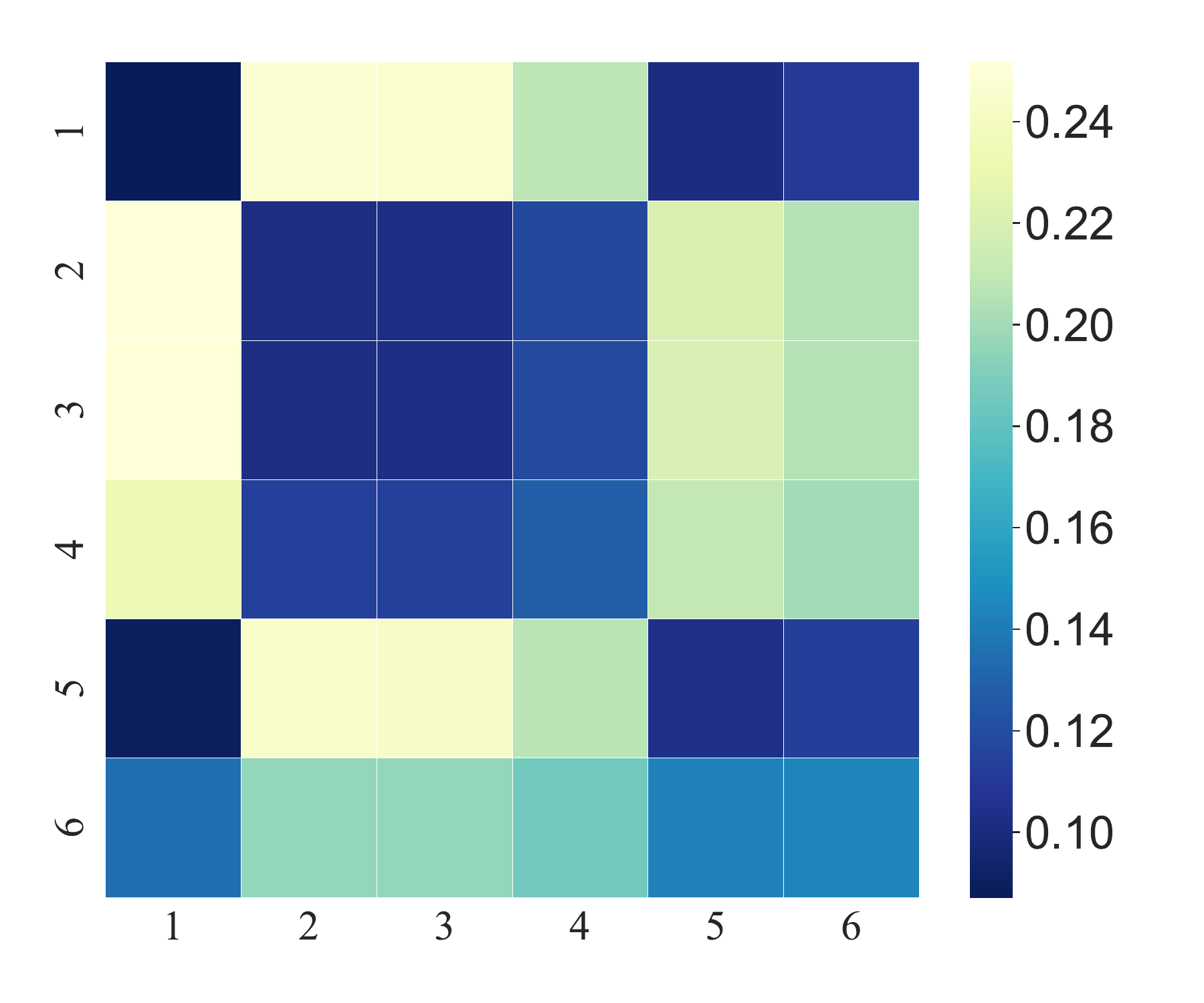} 
		\subcaption{Character Group}
		\label{fig:1_char_matrix}
	\end{minipage}
	\begin{minipage}[b]{0.45\textwidth}
		\includegraphics[width=\textwidth]{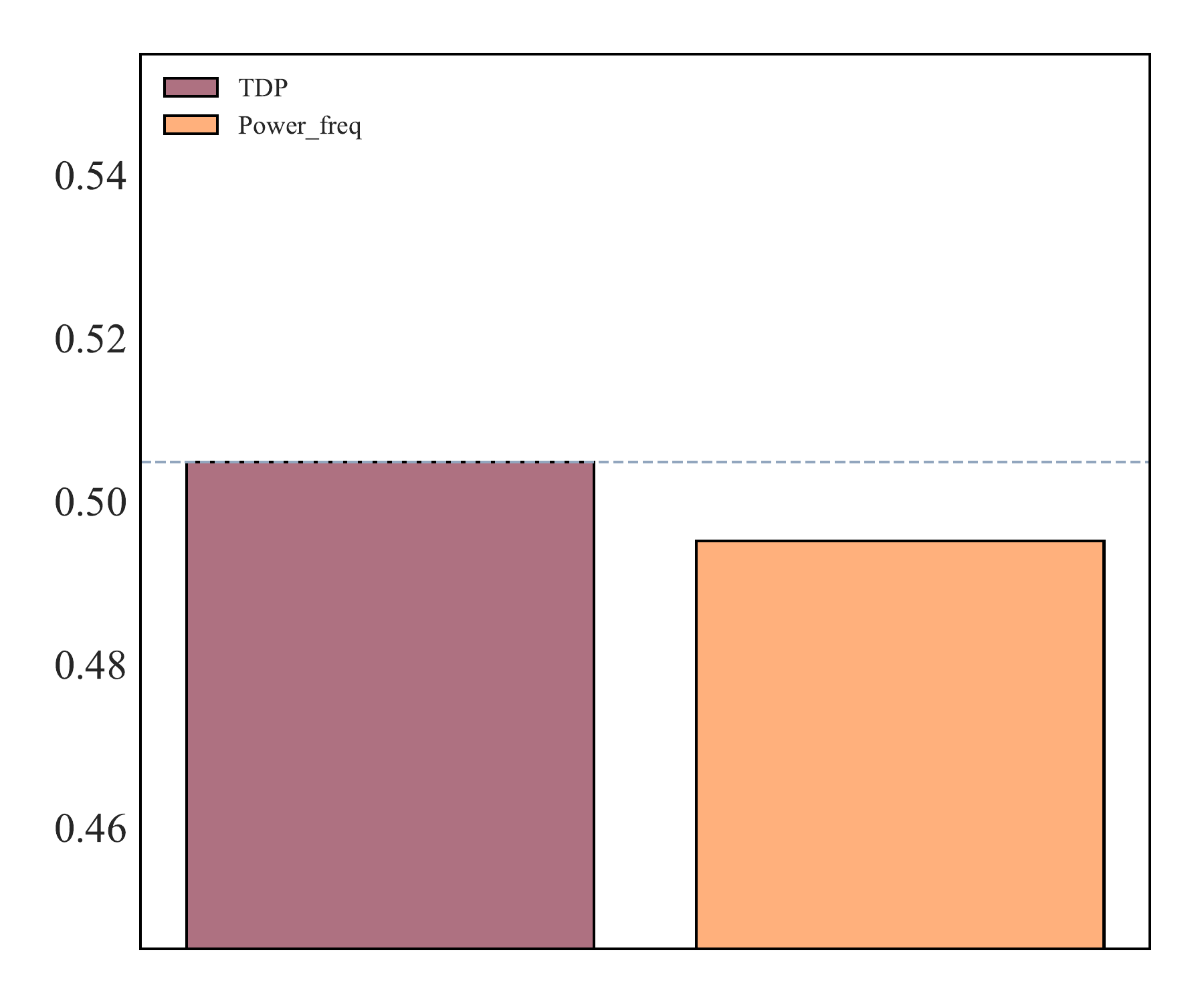}
		\subcaption{Other Group}
		\label{fig:1_system_bar}
	\end{minipage}
	\begin{minipage}[b]{0.45\textwidth}
		\includegraphics[width=\textwidth]{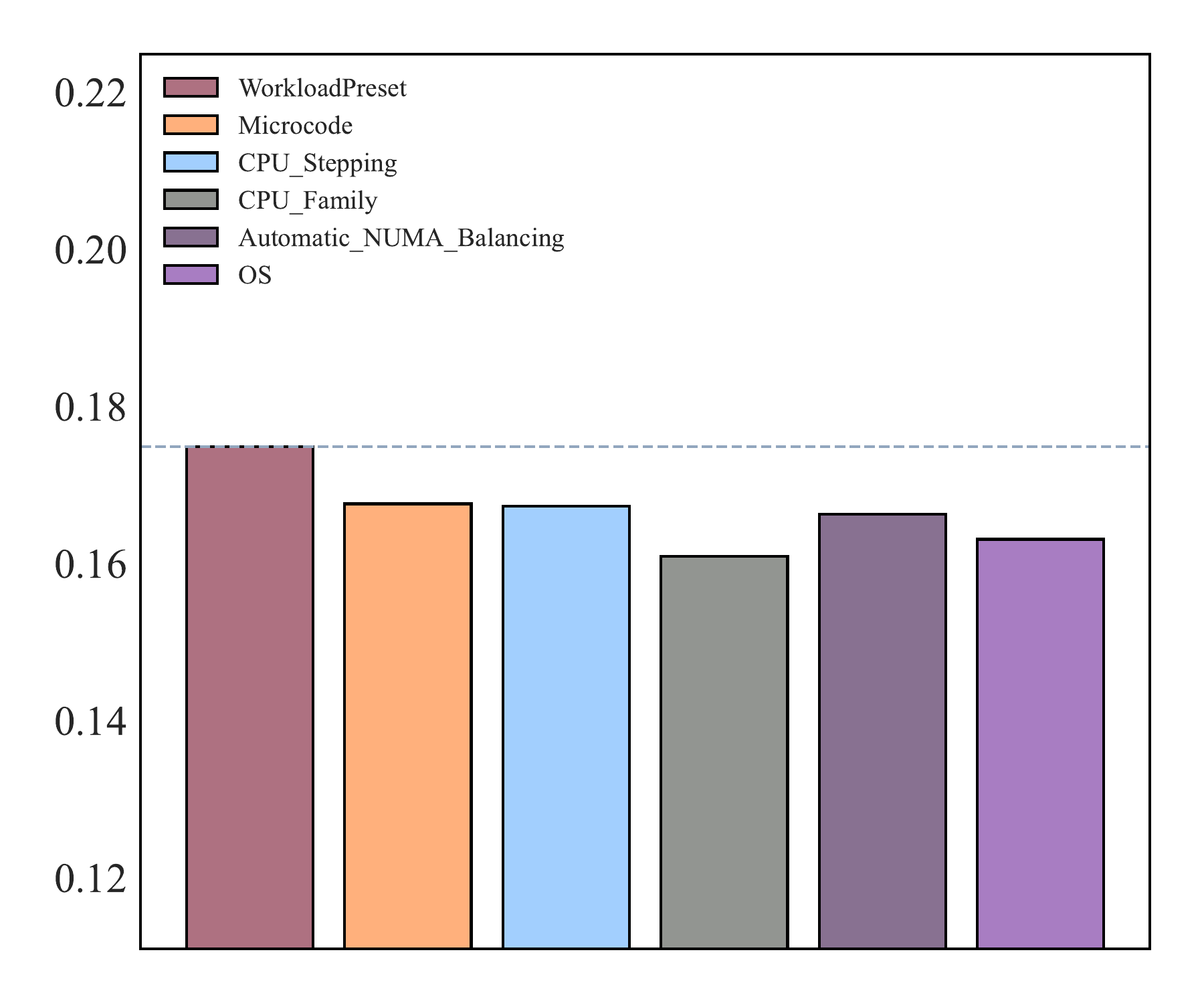}
		\subcaption{Character Group}
		\label{fig:1_char_bar}
	\end{minipage}
	\caption{Attention Matrix and Aggregated Feature Importance Visualization in Other and Character Characteristics Intra-groups of the First Head for a SPEC CPU 2017 FP Data Sample}
	\label{fig:1_0_attention}
\end{figure*}

\begin{figure*}[htbp]
	\begin{minipage}[b]{0.45\textwidth}
		\includegraphics[width= \textwidth]{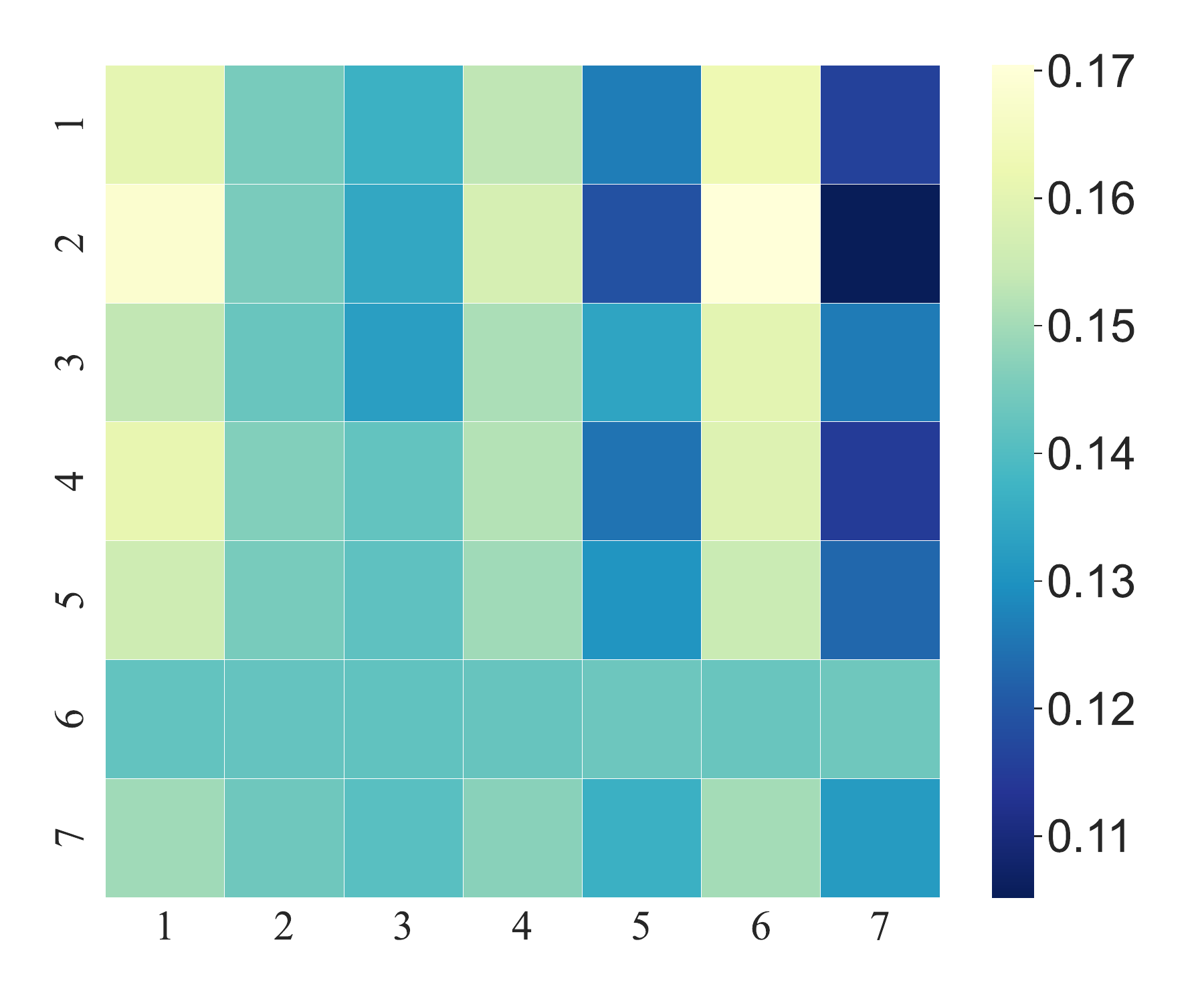} 
		\subcaption{Memory Group}
		\label{fig:1_mem_matrix}
	\end{minipage}
	\begin{minipage}[b]{0.45\textwidth}
		\includegraphics[width=\textwidth]{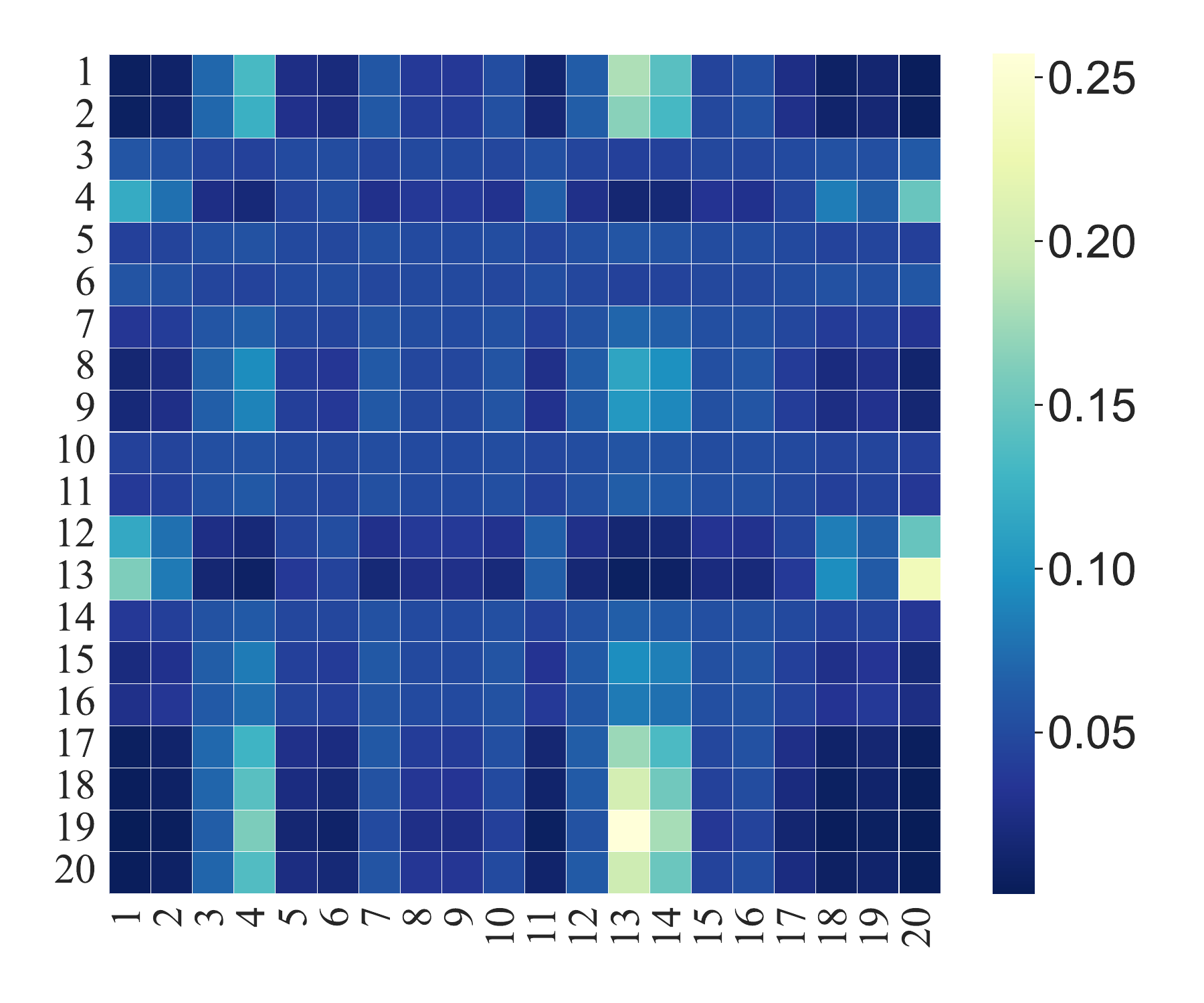} 
		\subcaption{CPU Group}
		\label{fig:1_cpu_matrix}
	\end{minipage}
	\begin{minipage}[b]{0.49\textwidth}
		\includegraphics[width=\textwidth]{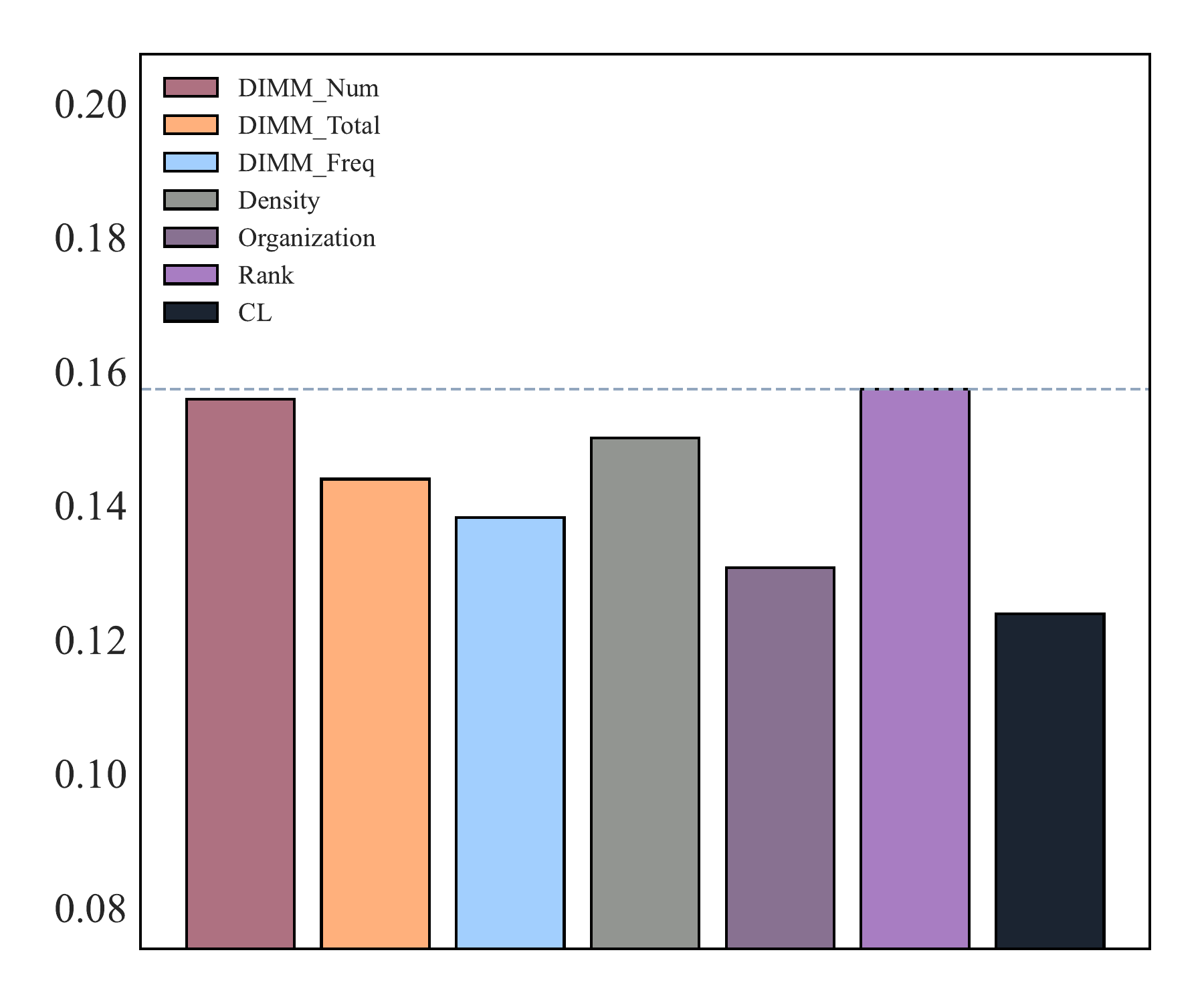} 
		\subcaption{Memory Group}
		\label{fig:1_mem_bar}
		\end{minipage}
		\begin{minipage}[b]{0.49\textwidth}
			\includegraphics[width=\textwidth]{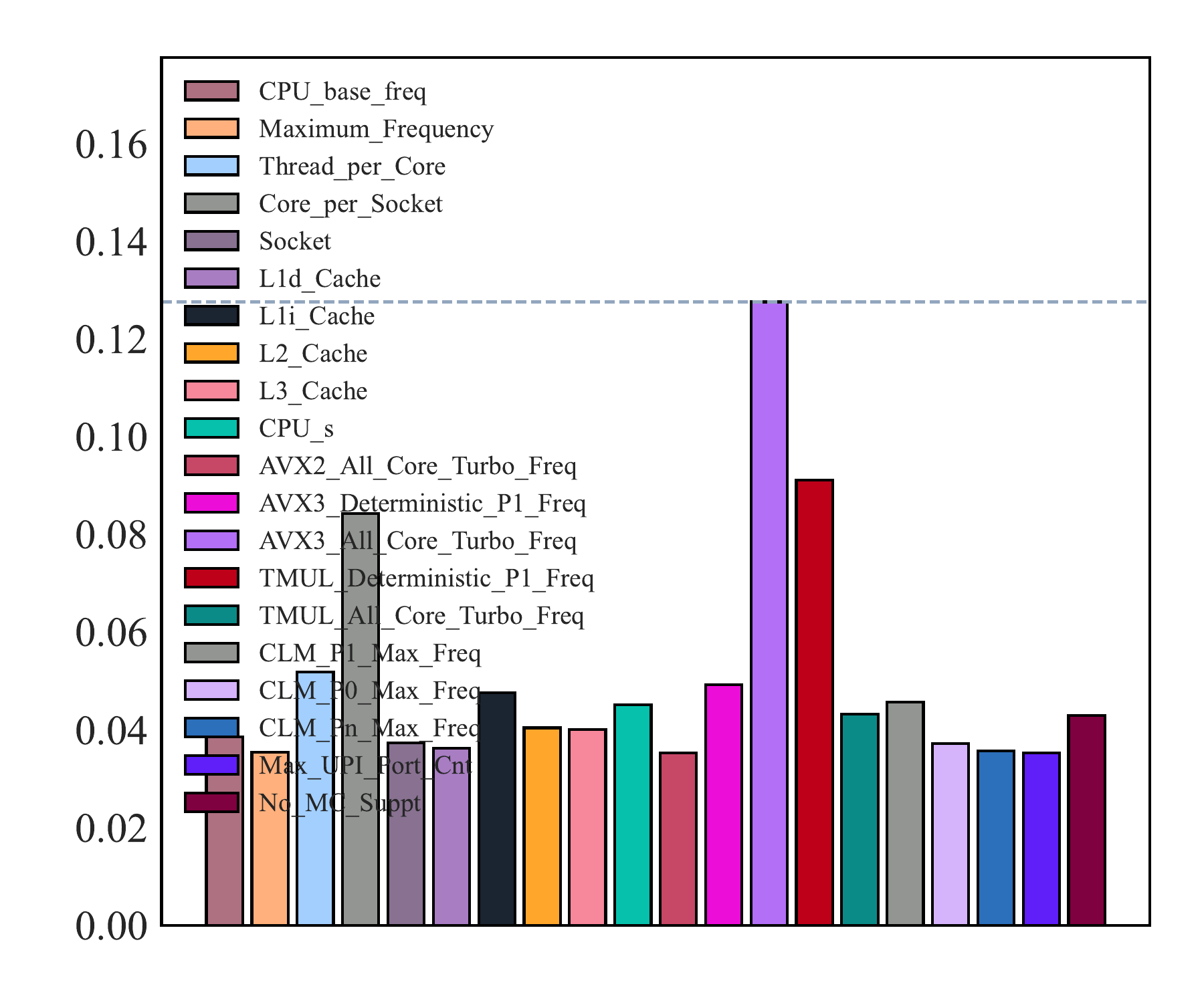}
			\subcaption{CPU Group}
			\label{fig:1_cpu_bar}
		\end{minipage}
		\caption{Attention Matrix and Aggregated Feature Importance Visualization in Memory and Processor Characteristics Intra-groups of the First Head for a SPEC CPU 2017 FP Data Sample}
		\label{fig:1_1_attention}
	\end{figure*}

\begin{figure*}[htbp]
	\centering
\begin{minipage}[b]{0.45\textwidth}
	\includegraphics[width=\textwidth]{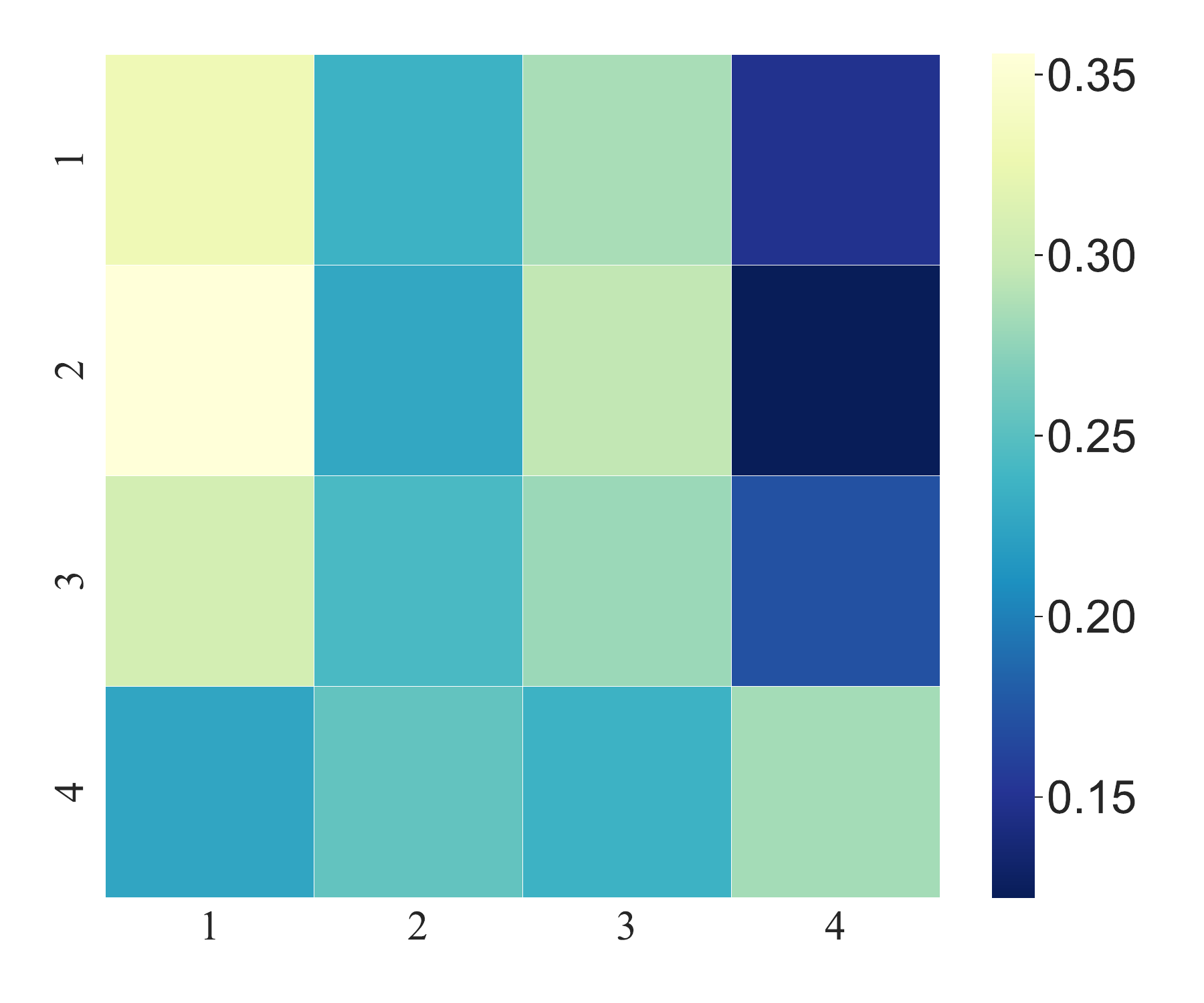}
	\label{fig:1_cross_matrix}
\end{minipage}
\begin{minipage}[b]{0.45\textwidth}
	\raisebox{0.4cm}{
	\includegraphics[width=\textwidth]{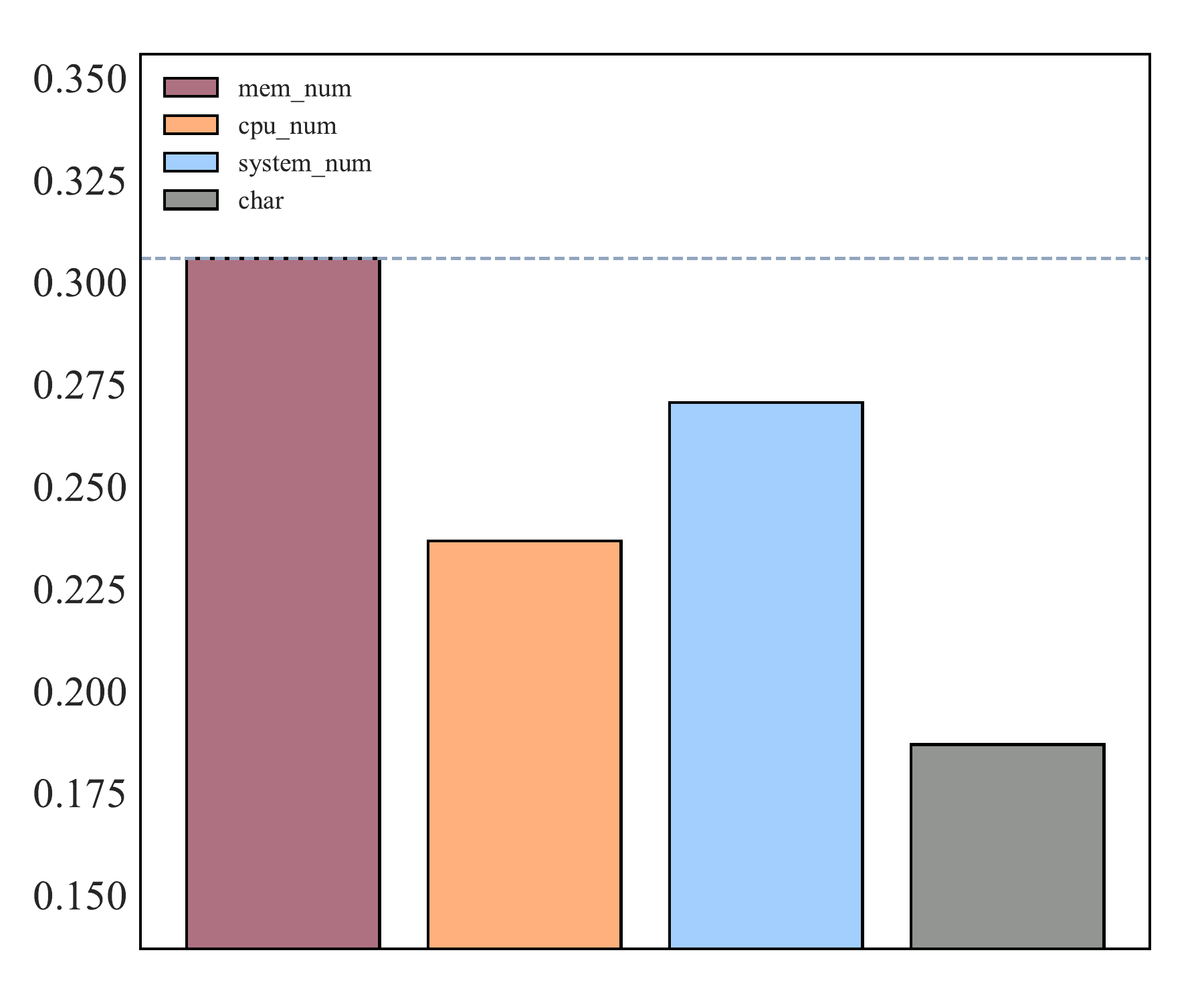}}
	\label{fig:1_cross_bar}
\end{minipage}
\caption{Attention Matrix and Aggregated Feature Importance Visualization in the Inter-group of the First Head for a SPEC CPU 2017 FP Data Sample}
\label{fig:1_attention}
\end{figure*}

The heatmaps depicted in Figure.\ref{fig:1_0_attention} - Figure.\ref{fig:1_attention} visualize the attention matrices for both intra-group and inter-group attention mechanisms within head 1. The color intensity in each cell of the heatmap corresponds to the attention weight assigned to each feature group. Lighter shades represent higher attention weights, whereas darker shades correspond to lower weights. This visualization helps to understand which hardware features of the input data are deemed more important by each head, thereby highlighting the model focus areas.

Figure.\ref{fig:1_0_attention} provides a comprehensive visualization of the attention matrix and aggregated feature importance for both the "Other Group" and "Character group" in the model trained with "SPECrate2017 FP" data. The attention matrix for the "Character Group" in Figure.\ref{fig:1_char_matrix} illustrates the attention weights among 6 features, revealing multiple strong interactions, which underscores the importance of capturing these interactions for accurate predictions. The aggregated feature importance for the "Other Group" in Figure.\ref{fig:1_system_bar} shows that "TDP" has a slightly higher importance compared to "Power\_freq," indicating that "TDP" plays a more critical role in the prediction. For the "Character Group" in igure.\ref{fig:1_char_bar}, the importance of features is more evenly distributed, with "Preset," "Microcode," and "CPU\_Stepping" exhibiting higher importance, while "CPU\_Family" and "OS" have relatively lower importance. This distribution suggests that multiple features within this group contribute significantly to the model performance, and their interactions are essential for capturing the underlying patterns in the data.

Figure.\ref{fig:1_1_attention} provides a comprehensive visualization of the attention matrix and aggregated feature importance for the "Memory Group" and "CPU Group" within the "SPECrate2017 FP" model. The attention matrix for the "Memory Group" in Figure.\ref{fig:1_mem_matrix} illustrates significant interactions between the seven features, crucial for understanding their collective influence on the model's instance-wise predictions. In contrast, the "CPU Group" in Figure.\ref{fig:1_cpu_matrix} shows multiple strong interactions among twenty features, underscoring the importance of capturing these interactions for accurate predictions. The aggregated feature importance for the "Memory Group" in Figure.\ref{fig:1_mem_bar} reveals that "DIMM\_rank", "Density", and "DIMM\_Num" are the most critical feature, followed by "DIMM\_Total" and "DIMM\_Freq," while "Organization" and "CL" have lower importance. For the "CPU Group" in Figure.\ref{fig:1_cpu_bar}, "AVX3\_TurboFreq" exhibits the highest importance, with other features like "TMUL\_P1Freq" and "Core\_per\_Socket" also contributing significantly.

Figure.\ref{fig:1_attention} provides a comprehensive visualization of the inter-group attention matrix for the first head in the “SPECrate2017 FP” model. This inter-group attention highlights dependencies across different feature groups, enhancing our understanding of feature interactions in model predictions. Among inter-group attention mechanisms, "Memory group", "Other Group" (including "TDP") and "CPU group" receive the most substantial attention weights overall. This indicates that features related to both memory and CPU have significant effects on performance compared to other groups. In contrast, "Char Group" features show minimal importance.

The utilization of both intra-group and inter-group attention mechanisms significantly enhances the ability of the model to discern and prioritize hardware configurations from complex datasets. Intra-group attention focuses on fine-grained details within subsets of data, providing a deeper understanding of internal dynamics and relationships. Inter-group attention captures dependencies between different feature groups, offering a broader perspective on feature interactions. The combined use of these mechanisms enhances model interpretability by clarifying which features and feature groups exert the most influence in the decision-making process.

\section{Conclusion}\label{sec13}
We propose a comprehensive benchmark dataset, PerfCastDB, specifically designed for CPU performance prediction tasks. Additionally, we introduce the NCPP as a baseline model. NCPP effectively categorizes and models hardware characteristics to enhance prediction accuracy. Extensive experiments demonstrate that NCPP achieves superior performance compared to traditional approaches, validating its effectiveness. We hope that our work provides a robust foundation for future research in CPU performance prediction.

For future work, we aim to further optimize the NCPP model structure by exploring greater complexity to enhance predictive accuracy and interpretability. We also plan to enrich the dataset by including a wider variety of CPUs and application scenarios to validate the model on larger and more complex data.

\backmatter

\bmhead{Acknowledgements}
I would like to express my sincere gratitude for the support and contributions from my colleagues and organization in the preparation of this paper. Special thanks to Zhang Jiajia for providing the original SPR data for the time period mentioned in this paper. My gratitude also goes to Benson and Xiaoping Zhou for sharing their background knowledge on the dataset hardware characteristics. And I appreciate Xiaofei, Lei Rong, and Wenchao He for their valuable suggestions on optimizing the code style during the project. Special appreciation is extended to Zhongbin, for his support throughout this project.

I would also like to extend my heartfelt gratitude to Jingwei Cai from Tsinghua University. Despite the constraints of not disclosing any dataset information, his invaluable assistance in revising this paper and his insightful suggestions have significantly contributed to the improvement of this work.

Finally, I would like to thank Intel for their financial support and resources, which made this research possible.



\section*{Declarations}

\begin{itemize}
\item Funding:This research was supported in part by Intel Corporation. The company may cover the publication fee for this manuscript.
\item Conflict of interest/Competing interests: The authors declare that there are no known competing financial interests or personal relationships that could have appeared to influence the work reported in this paper. Although the author is an employee of Intel Corporation, the development of the model and the preparation of this manuscript were conducted independently of the company's direct business interests.
\item Ethics approval and consent to participate: Not applicable. The dataset used for this research is proprietary to Intel Corporation and does not involve any human subjects or sensitive personal data that would require ethical approval. All procedures followed were in accordance with corporate policies and legal requirements.
\item Consent for publication: Not applicable. This manuscript does not contain any individual person’s data
\item Data availability: The dataset utilized for developing the CPU performance prediction model is proprietary to Intel Corporation and cannot be made publicly available due to confidentiality agreements and the sensitive nature of the business. The use of the data strictly adhered to the company's data protection policies.
\item Materials availability: Not applicable. Although the model training process involved the use of computational resources, no physical materials were utilized in this research..
\item Code availability: The code for the CPU performance prediction model developed in this study is publicly available. The code can be accessed on GitHub: [\url{https://github.com/xiaoman-liu/NCPP}].
\item The author was solely responsible for the procesing of the data, development of the model, conducting experiments, and the drafting and revising of this manuscript. Colleagues within Intel Corporation provided insights by sharing background information on features and offering their perspectives on the features. However, the author independently carried out the whole work.
\end{itemize}

\noindent
\bibliography{refs}







\end{document}